\documentclass[12pt]{article}
\usepackage{amsmath,amssymb} 
\usepackage{subcaption}   
\usepackage{amsfonts}
\usepackage{mathrsfs}
\usepackage{enumerate}
\usepackage{ytableau}
\usepackage{hyperref}
\numberwithin{equation}{section}
\usepackage{cite}
\usepackage{bm}
\usepackage{tikz}
\usepackage{blkarray}  
\usepackage{xcolor}

\begin{document}

\def\im{\text{i}}
\def\be{\begin{equation}}
\def\ee{\end{equation}}
\def\ba{\begin{array}}
\def\ea{\end{array}}
\def\bd{\begin{displaymath}}
\def\ed{\end{displaymath}}
\def\eg{{\it e.g.~}}
\def\ie{{\it i.e.~}}
\def\Tr{{\rm Tr}}
\def\tr{{\rm tr}}
\def\>{\rangle}
\def\<{\langle}
\def\a{\alpha}
\def\b{\beta}
\def\c{\chi}
\def\del{\delta}
\def\e{\epsilon}
\def\f{\phi}
\def\vf{\varphi}
\def\tvf{\tilde{\varphi}}
\def\g{\gamma}
\def\h{\eta}
\def\j{\psi}
\def\k{\kappa}
\def\l{\lambda}
\def\m{\mu}
\def\n{\nu}
\def\w{\omega}
\def\p{\pi}
\def\q{\theta}
\def\r{\rho}
\def\s{\sigma}
\def\t{\tau}
\def\u{\upsilon}
\def\x{\xi}
\def\z{\zeta}
\def\D{\Delta}
\def\F{\Phi}
\def\G{\Gamma}
\def\J{\Psi}
\def\L{\Lambda}
\def\W{\Omega}
\def\P{\Pi}
\def\Q{\Theta}
\def\S{\Sigma}
\def\U{\Upsilon}
\def\X{\Xi}
\def\nab{\nabla}
\def\pa{\partial}
\newcommand{\lra}{\leftrightarrow}

\newcommand{\bc}{{\mathbb{C}}}
\newcommand{\br}{{\mathbb{R}}}
\newcommand{\bz}{{\mathbb{Z}}}
\newcommand{\bp}{{\mathbb{P}}}

\def\nn{\nonumber \\}

\newcommand{\red}{\textcolor[RGB]{255,0,0}}
\newcommand{\blue}{\textcolor[RGB]{0,0,255}}
\newcommand{\green}{\textcolor[RGB]{0,255,0}}
\newcommand{\cyan}{\textcolor[RGB]{0,255,255}}
\newcommand{\magenta}{\textcolor[RGB]{255,0,255}}
\newcommand{\yellow}{\textcolor[RGB]{255,255,0}}
\newcommand{\sky}{\textcolor[RGB]{135, 206, 235}}
\newcommand{\orange}{\textcolor[RGB]{255, 127, 0}}
\def\d{\operatorname{d}}
\def\co{\mathcal{O}}
\def\arctanh{\operatorname{arctanh}}
\def\sech{\operatorname{sech}}
\title{\textbf{Comments on holographic spread complexity}}
\vspace{14mm}
\author{Zhehan Li \footnote{lizhehan@sxu.edu.cn} and Jia Tian \footnote{wukongjiaozi@ucas.ac.cn}
}
\date{}
\maketitle

\begin{center}
	{\it State Key Laboratory of Quantum Optics and Quantum Optics Devices, Institute of Theoretical Physics, Shanxi University, Taiyuan 030006, P.~R.~China
	}
\vspace{10mm}
\end{center}

\makeatletter
\def\blfootnote{\xdef\@thefnmark{}\@footnotetext}  
\makeatother

\begin{abstract}
We revisit the holographic proposal relating the growth rate of spread complexity to the radial momentum of a bulk probe, aiming to identify its underlying assumptions and clarify how classical probe dynamics emerges from quantum dynamics.
By quantizing AdS probes directly and using the extrapolation dictionary, we provide a more general derivation of the proposal within the probe approximation for the specified class of states. In particular, we discuss a probe analogue of the classical complexity observables appearing in the ``complexity=anything'' (CAny) framework and examine how such an observable can be quantized.
Since probe dynamics is not intrinsically tied to the AdS/CFT correspondence, we also examine the proposal in flat and de Sitter spacetimes. In flat spacetime, the physical Hamiltonian does not preserve the coherent-state structure, and spread complexity measures the dispersive broadening of the wave packet rather than classical momentum. In de Sitter space, generic semiclassical wave packets in the static-patch energy representation show the same dispersive behavior. By contrast, coherent states adapted to the principal-series representation can exhibit exponential growth of spread complexity, with a growth rate that has the same time dependence as the classical radial momentum. These examples indicate that a semiclassical limit alone is not sufficient for a momentum--complexity relation.
\end{abstract}

\baselineskip 18pt
\newpage

\tableofcontents

\section{Introduction}
One of the most striking features of the AdS/CFT correspondence \cite{Maldacena:1997re,Witten:1998qj,Gubser:1998bc} is that certain quantum observables admit simple geometric descriptions. The Ryu--Takayanagi formula and its covariant extension provide a paradigmatic example, expressing boundary entanglement entropy as the area of an extremal surface in the bulk \cite{Ryu:2006bv,Ryu:2006ef,Hubeny:2007xt}. Beyond such geometric descriptions of entropy, holography also suggests that certain aspects of quantum evolution may be encoded in classical gravitational dynamics. For example, the growth of entanglement entropy after a global quench \cite{Calabrese:2004eu,Calabrese:2005in} is captured by the growth of the black hole interior in the dual geometry \cite{Hartman:2013qma}. However, this example also reveals the limitations of entanglement as a probe of late-time dynamics: coarse-grained entanglement entropy saturates after thermalization, while the quantum state continues to evolve and the interior of a classical black hole keeps growing.

Thus, while entanglement entropy captures important aspects of early-time relaxation, it does not provide a sufficiently fine probe of the continued evolution of the quantum state or of the growing black hole interior. This motivates the introduction of more refined diagnostics of quantum evolution.

Quantum complexity provides a natural candidate \cite{Susskind:2014rva}, since its saturation time is expected to be exponentially longer than that of entanglement entropy. The CAny proposal \cite{Belin:2021bga,Belin:2022xmt}, building on the original holographic prescriptions for complexity \cite{Susskind:2014rva,Brown:2015bva,Couch:2016exn}, sharpens this question by suggesting that infinitely many classical gravitational observables can track post-thermalization dynamics. It therefore calls for identifying their boundary quantum counterparts. To make such conjectures concrete, and ultimately testable, one needs computable notions of quantum complexity \cite{Chapman:2017rqy}. Krylov complexity \cite{Parker:2018yvk}, recently reviewed in \cite{Nandy:2024htc,Baiguera:2025dkc,Rabinovici:2025otw}, provides an operator-based diagnostic, while spread complexity \cite{Balasubramanian:2022tpr} gives a closely related notion for states. The distinctions and possible connections between Krylov, circuit, and Nielsen complexities are discussed in \cite{Aguilar-Gutierrez:2023krylov,Craps:2025krylov}. 
In lower-dimensional models, especially in the DSSYK/JT setting \cite{Berkooz:2018jqr,Berkooz:2018qkz,Garcia-Garcia:2017pzl,Jackiw:1984je,Teitelboim:1983ux}, spread complexity has been shown to match the renormalized length of the wormhole connecting the two boundaries in suitable states and regimes \cite{Lin:2022rbf,Rabinovici:2023yex,Aguilar-Gutierrez:2025pqp}. Related operator K-complexity constructions have been studied in \cite{Jian:2020qpp,Ambrosini:2024sre,Aguilar-Gutierrez:2025pqp}. Further developments can be found in \cite{Erdmenger:2023wjg,Nandy:2024tridiagonal,Ambrosini:2025hvo,Heller:2024ldz,Balasubramanian:2024lqk,Xu:2024gfm,Aguilar-Gutierrez:2025evolution,Aguilar-Gutierrez:2025symmetry,Alfinito:2026cky,Aguilar-Gutierrez:2026nmd,Aguilar-Gutierrez:2026ogo,Aguilar-Gutierrez:2026probing,Fu:2025kkh,Heller:2025ddj,Miyaji:2025yvm} and references therein. Extending these precise relations to higher-dimensional theories, however, remains challenging.

A different, but closely related, way in which complexity is expected to encode bulk dynamics concerns the motion of infalling probes. The proposed relation takes the form
\be
\frac{d C_{\mathcal{O}}}{dt}\propto p_{\text{infalling}}, \label{dC}
\ee
as discussed in \cite{Susskind:2014jwa,Susskind:2018tei,Susskind:2019ddc,Brown:2018kvn,Susskind:2020gnl,Magan:2018nmu,Barbon:2020uux,Barbon:2020olv,Barbon:2019tuq,Lin:2019kpf,Lin:2019qwu,Ageev:2018msv,Ageev:2018nye}. Here $C_{\mathcal{O}}$ denotes the operator size, or operator complexity, and $p_{\text{infalling}}$ is the momentum of a freely falling bulk particle. A precursor of this idea appeared in \cite{Balasubramanian:1998de}, where the dynamics of a bulk probe was related to the spreading of localized fields in the boundary theory. The relation \eqref{dC} was verified in the JT/SYK duality \cite{Kitaev:2015,Lin:2019qwu}, building on earlier studies of operator growth in the SYK model \cite{Sachdev:1992fk,Qi:2018bje,Roberts:2018mnp}. More recently, analogous relations have been examined in holographic CFT setups \cite{Caputa:2024sux}; see also \cite{Fan:2024iop,He:2024pox,Aguilar-Gutierrez:2025kmw,Li:2025fqz}. Related holographic proposals for spread complexity have attracted considerable attention \cite{Das:2024tnw,Caputa:2025dep,Fatemiabhari:2025cyy,Fatemiabhari:2025usn,Roychowdhury:2026eds,Fatemiabhari:2026rob,Zoakos:2026obl,Nastase:2026lhz,Roychowdhury:2026sgg,Alfinito:2026vah,Graef:2026pzv,Chatzis:2026ekd,BitaghsirFadafan:2026lek,Basu:2026aky}.

In this work, we revisit this proposal with two aims. The first is to make explicit the assumptions on which it rests and to determine the extent to which it depends on the AdS/CFT correspondence. The second is to clarify how the classical probe dynamics emerges from the underlying quantum dynamics. Once these points are fixed, several generalizations become natural. In contrast to much of the recent literature, we follow the viewpoint of \cite{Balasubramanian:1998de} and focus directly on the probe dynamics. By quantizing the probe system and using the standard AdS/CFT extrapolation dictionary, we identify the conditions under which spread complexity tracks classical radial momentum.

\subsection{A brief review of the proposal}
\label{sec_intro}
We begin by reviewing the proposal introduced in \cite{Caputa:2024sux} and further elaborated in \cite{Li:2025fqz} for global AdS$_3$,
\be
ds^2=-\cosh^2\rho\,dt^2+d\rho^2+\sinh^2\rho\,d\phi^2,
\qquad
\phi\sim \phi+2\pi .
\ee
After the Wick rotation $t\to-\im t_E$, the boundary cylinder is mapped to the complex plane by the exponentiation
\be
z=e^{t_E+\im\phi},
\qquad
\bar z=e^{t_E-\im\phi}.
\ee
Quantum states are prepared by the Euclidean path integral on the unit disk. We consider the initial state associated with the insertion of a scalar primary operator of dimension $\Delta$,
\be
|\psi_0\rangle
=
\frac{1}{\mathcal N}\co(z_0,\bar z_0)|\Omega\rangle,
\qquad
|z_0|<1 ,
\label{adsstate}
\ee
where $|\Omega\rangle$ denotes the vacuum.  Since the motion is effectively radial, rotational symmetry in the $\phi$ direction allows us to take $z_0$ to be real, or equivalently $\phi=0$. The CFT Hamiltonian generating Lorentzian evolution in global time is
\be
H_{\rm CFT}=L_0+\bar L_0\equiv l_0,
\ee
and hence
\be
|\psi(t)\rangle
=
e^{-\im l_0t}|\psi_0\rangle .
\ee
Here $L_0,L_{\pm1}$ and $\bar L_0,\bar L_{\pm1}$ denote the generators of the holomorphic and antiholomorphic global conformal sectors associated with the $z$ and $\bar z$ coordinates, respectively. This dynamics is usually referred to as a local operator quench \cite{Nozaki:2014hna,Caputa:2014vaa,Nozaki:2013wia}.

In the semiclassical regime
\be
1\ll\Delta\ll \frac{1}{G_N},
\ee
the initial state $|\psi_0\rangle$ was proposed to be dual to a massive particle located at $\rho_0=2\arctanh z_0$, while $|\psi(t)\rangle$ is dual to its free fall in global AdS. The particle follows the timelike geodesic
\be
\tanh\rho=\tanh\rho_0\cos t,
\qquad
t\geq0 .
\label{geodesic}
\ee
Thus the CFT time evolution generated by $l_0$ is identified with the physical geodesic motion of the bulk particle.

For the computation of spread complexity, the key observation of \cite{Li:2025fqz} is that the state
\be
|\psi_0\rangle=\co(z_0)|\Omega\rangle
\ee
is itself a highest-weight state, but with respect to a transformed set of generators
\be
\mathcal L_n
=
e^{\im Q}l_ne^{-\im Q},
\qquad l_n=L_n+\bar{L}_n,\qquad 
n=0,\pm1,
\ee
where the $SL(2,\mathbb{R})$ transformation generated by $Q$ maps the origin to $z_0$. For the real insertion
\be
z_0=\tanh\frac{\rho_0}{2},
\label{z0rho}
\ee
one finds
\be
\mathcal L_0
=
\cosh\rho_0\,l_0
-
\sinh\rho_0\,\frac{l_{+1}+l_{-1}}{2}.\label{cl0}
\ee
In this adapted representation, the state $|\psi_0\rangle$ becomes a highest-weight state,
\be
\mathcal L_0|\psi_0\rangle=\Delta|\psi_0\rangle,
\qquad
\mathcal L_{+1}|\psi_0\rangle=0 .
\ee
Accordingly, the Krylov basis generated from the initial state $|\psi_0\rangle$ can be identified with the corresponding highest-weight module,
\be
|K_n\rangle
=
|\Delta,n\rangle_{z_0}
=
\sqrt{\frac{\Gamma(2\Delta)}{n!\Gamma(2\Delta+n)}}\,
\mathcal L_{-1}^{\,n}|\psi_0\rangle .
\ee
So the Krylov number operator is
\be
\hat n=\mathcal L_0-\Delta,
\ee
and the spread complexity and its growth rate are then
\begin{eqnarray}
C(t)
&=&
\langle\psi(t)|\mathcal L_0|\psi(t)\rangle
-\Delta
=
2\Delta \sinh^2\rho_0\sin^2\frac{t}{2}
=
\frac{8\Delta z_0^2}{(1-z_0^2)^2}\sin^2\frac{t}{2},
\label{ckt}
\\
\frac{dC(t)}{dt}
&=&
\im \langle\psi(t)|[l_0,\mathcal L_0]|\psi(t)\rangle =\frac{8\Delta z_0^2}{(1-z_0^2)^2}\sin t.
\end{eqnarray}
Since $C(t)$, and likewise $\dot C(t)$, is a linear combination of the expectation values of $l_0,l_{\pm1}$, the extrapolation dictionary permits a direct bulk evaluation. The global conformal generators lift to Killing vector fields in AdS. Let $\hat l_n$ denote the Killing vector associated with $l_n$, whose action on the bulk scalar field is given by
\be
[l_n,\hat{\phi}(0)]
=
\im \hat{l}_n^\mu \partial_\mu\hat{\phi}.
\ee
The corresponding expectation values can then be written as
\be
\langle \Omega|\hat{\phi}(0)e^{\im H t}  l_ne^{-\im H t}\hat\phi(0)|\Omega\rangle_{\text{AdS}}
=
\im \hat{l}_n^\mu(t) \langle\phi|\partial_\mu |\phi\rangle_{\text{AdS}},
\qquad
\hat{l}_n(t)\equiv \widehat{e^{\im H t}  l_ne^{-\im H t}} .
\label{bulkexpect}
\ee
In the semiclassical regime, ${p}_\mu\equiv \im \langle\phi|\partial_\mu |\phi\rangle_{\text{AdS}}$ reduces to the momentum of the massive particle. Thus the CFT expectation value of $l_n$ is mapped to the projected bulk momentum,
\be
\langle l_n(t)\rangle_{\rm CFT}
\quad\longleftrightarrow\quad
\hat l_n^\mu(t) p_\mu\Big|_{\rm geodesic},
\label{dictionary}
\ee
where $p_\mu$ is the particle momentum. Using the dictionary \eqref{dictionary} and the relation \eqref{cl0}, one obtains the bulk interpretation of the spread complexity and its growth rate:
\begin{eqnarray}
C(t)+\Delta
&=&
\text{energy measured by a bulk observer},
\\
\dot C(t)
&=&
\text{radial momentum measured by the same observer}.
\end{eqnarray}
The observer is determined by the adapted Krylov basis, or equivalently by the $SL(2,\mathbb{R})$ transformation that maps the reference primary state at the origin to the insertion point $z_0$. With a suitable choice of radial coordinate, the growth rate matches the time dependence of the radial momentum.

The proposal raises several questions:
\begin{enumerate}
	\item The dictionary \eqref{dictionary} is not a derivation from first principles. Its precise regime of validity is therefore unclear.
	\item Can the construction be extended beyond the semiclassical regime? Ideally, the correspondence should first be formulated quantum mechanically, with the semiclassical limit taken only afterward.
	\item The derivation appears to rely strongly on $SL(2,\mathbb R)$ symmetry. How should the correspondence be modified in backgrounds without this symmetry?
	\item Are there other natural setups, corresponding for example to different initial states?
	\item Is holography necessary for spread complexity to admit a classical interpretation?
	\item What is the relation between this proposal and CAny?
	\item Can the proposal go beyond AdS spacetimes or free-particle dynamics?
\end{enumerate}
We will address some of these points in what follows. Several further important issues remain:
\begin{enumerate}
	\item How is the picture modified by backreaction?
	\item What happens for other classical probes, such as strings or branes?
	\item Can spread complexity exhibit a switchback effect?
	\item What new insights can spread complexity provide into holography?
\end{enumerate}
We will not address these questions in this paper, leaving them for future work.

 The rest of the paper is organized as follows. In Section \ref{sec_revisit}, we revisit the spread-complexity proposal in global AdS. We first derive the spread complexity directly from the bulk scalar theory, then take the semiclassical limit and show how the classical particle trajectory emerges. We also quantize a free massive particle in AdS$_2$ and identify the corresponding complexity observable on the particle phase space. In Section \ref{sec_flat}, we study a free particle in flat spacetime. This example separates coherent motion from dispersive spreading and shows that, for the physical Hamiltonian, spread complexity measures the broadening of the wave packet rather than its classical momentum. In Section \ref{sec_dS}, we discuss the de Sitter case. We compare semiclassical wave packets in the static-patch energy representation with states adapted to the principal-series representation, and show when the growth rate of spread complexity can be related to radial momentum. We also comment on the relation to JT gravity and DSSYK. We conclude in Section \ref{sec_summary} with a summary and several open directions. Appendix \ref{appendix_adsd} contains the higher-dimensional AdS calculation of the spectral weights.

\section{Revisiting spread complexity in AdS}
\label{sec_revisit}
We now revisit the calculation of spread complexity from the bulk point of view. We use only two basic ingredients of AdS/CFT: the extrapolation dictionary and the identification of the boundary Hamiltonian with the bulk Hamiltonian. Since the dynamics is effectively confined to AdS$_2$, we restrict attention to global AdS$_2$. In Appendix \ref{appendix_adsd}, we carry out a similar calculation in global AdS$_{d+1}$ spacetime and obtain essentially the same result.
\subsection{Free massive scalar field on global \texorpdfstring{AdS$_2$}{AdS2}}
We begin with some standard facts about the boundary/bulk correspondence for probe dynamics
\cite{Balasubramanian:1998sn,Balasubramanian:1998de,Hamilton:2005ju,Hamilton:2006az}. In particular, it was proposed in \cite{Balasubramanian:1998de} that ``strings or particles move in AdS spacetime to reduce gravitational potential energy; this is dual in the boundary theory to the spreading of localized field distributions to reduce gradient energy.'' The holographic interpretation of spread complexity may be viewed as a concrete realization of this idea.

Consider a real massive scalar field of mass $m$ on global AdS$_2$ with metric
\be
ds^2=-\cosh^2\rho\,dt^2+d\rho^2,
\qquad
-\infty<\rho<\infty ,
\ee
whose action is given by
\be
S[\phi]
=
-\frac{1}{2}\int_M d^2x\,\sqrt{-g}\,
\left(
g^{\mu\nu}\partial_\mu\phi\partial_\nu\phi
+
m^2\phi^2
\right).
\ee
For a generic mass, the scalar field admits the mode expansion
\begin{eqnarray}
\phi
=
\sum_{n=0}^\infty
\left(
a_n u_n+a_n^\dagger u_n^*
\right),
\qquad
u_n
=
\mathcal{N}_n
\sech^\Delta \rho\,
C_n^\Delta(\tanh\rho)
e^{-\im\omega_n t},
\label{def_un}
\end{eqnarray}
where
\be
\Delta
=
\frac{1}{2}
\left(
1+\sqrt{1+4m^2}
\right),
\qquad
\omega_n=n+\Delta,
\qquad
\mathcal{N}_n
=
2^{\Delta-1}\Gamma(\Delta)
\sqrt{
\frac{\Gamma(n+1)}{\pi \Gamma(n+2\Delta)}
},
\ee
and $C_n^\Delta(x)$ is the Gegenbauer polynomial. Upon quantization, the coefficients $a_n$ and $a_n^\dagger$ become operators satisfying
\be
[\hat{a}_n,\hat{a}^\dagger_m]=\delta_{nm}.
\ee
The bulk Fock vacuum is defined by
\be
\hat a_n|\Omega\rangle_{\mathrm{bulk}}=0 .
\ee
The extrapolation dictionary gives the near-boundary behavior of a normalizable bulk field,
\be
\hat{\phi}(t,\rho)
\sim
(\cosh\rho)^{-\Delta}\mathcal{O}(t),
\qquad
\rho\to\infty .
\ee
Substituting the mode expansion into this relation yields the boundary mode expansion
\be
\mathcal{O}(t)
=
\sum_{n=0}^{\infty}
\left(
\beta_n e^{-\im\omega_n t}\hat a_n
+
\beta_n^* e^{\im\omega_n t}\hat a_n^\dagger
\right),
\label{bme}
\ee
where
\be
\beta_n
=
\frac{
2^{-\Delta}
\sqrt{\Gamma(n+2\Delta)}
}{
\Gamma\left(\Delta+\frac{1}{2}\right)
\sqrt{\Gamma(n+1)}
}.
\ee
Therefore the initial state
\be
|\psi_0\rangle
=
\frac{1}{\mathcal{N}_\co}
\co(z_0)|\Omega\rangle
\ee
is dual to the one-particle bulk state
\begin{eqnarray}
|\Psi_0\rangle
&=&
\frac{1}{\mathcal{N}_\co}
\sum_n
\left(
\beta_n z_0^{-n}\hat{a}_n
+
\beta_n^* z_0^n \hat{a}_n^\dagger
\right)
|\Omega\rangle_{\text{bulk}}
\nonumber\\
&=&
\frac{1}{\mathcal{N}_\co}
\sum_n
\beta_n^* z_0^n
\hat{a}_n^\dagger
|\Omega\rangle_{\text{bulk}}
=
\sum_n c_n
\hat a_n^\dagger
|\Omega\rangle_{\text{bulk}} .
\label{bulkinitial}
\end{eqnarray}
Here the normalization factor and the coefficients $c_n$ are
\begin{eqnarray}
\mathcal{N}_\co
&=&
(1-z_0^2)^{-\Delta}
2^{-\Delta}
\frac{\sqrt{\Gamma(2\Delta)}}{\Gamma\left(\Delta+\frac{1}{2}\right)},
\\
c_n
&=&
\frac{
z_0^n
\left(1-z_0^2\right)^{\Delta}
\sqrt{\Gamma(n+2\Delta)}
}{
\sqrt{\Gamma(2\Delta)}
\sqrt{\Gamma(n+1)}
}
=
z_0^n
\left(1-z_0^2\right)^{\Delta}
\sqrt{\frac{(2\Delta)_n}{n!}} .
\label{coe_cn}
\end{eqnarray}

\subsection{Spread complexity from the bulk}

Identifying the boundary Hamiltonian with the bulk Hamiltonian gives the time-evolved bulk state
\be
|\Psi(t)\rangle
=
\sum_n
c_n e^{-\im \omega_n t}
\hat{a}_n^\dagger|\Omega\rangle_{\text{bulk}} .
\ee
In terms of the bulk modes, the initial state $|\Psi_0\rangle\equiv|K_0\rangle$ does not make the $SL(2,\mathbb{R})$ symmetry manifest. As a result, it is not obvious how to implement the boundary construction of the Krylov basis directly in the bulk theory. Moreover, since that construction is closely tied to the underlying symmetry structure, it is not well suited for generalizations beyond the symmetric setting considered above.

Although one could in principle perform the Lanczos construction directly in the bulk Hilbert space, doing so is technically cumbersome. Instead, we adopt an equivalent approach that is more convenient for the present purpose and more readily generalizable. For the initial state \eqref{bulkinitial}, all Lanczos data are encoded in the spectral measure
\be
d\mu(E)
=
\sum_{n=0}^{\infty}
|c_n|^2\delta(E-\omega_n)dE .
\ee
Thus one may construct the orthogonal polynomials associated with $d\mu(E)$,
\be
\int d\mu(E)\,p_m(E)p_n(E)=\delta_{mn}.
\ee
Their three-term recursion relation,
\be
E p_n(E)
=
b_{n+1}p_{n+1}(E)+a_np_n(E)+b_np_{n-1}(E),
\ee
is precisely the Lanczos recursion relation. The relation between Krylov methods and orthogonal polynomials is standard; recent discussions include \cite{Muck:2022xfc,Balasubramanian:2025xkj}.

Using \eqref{coe_cn}, the spectral weight is
\be
w_n
=
|c_n|^2
=
q^n(1-q)^{2\Delta}
\frac{(2\Delta)_n}{n!},
\qquad
q\equiv z_0^2 .
\label{weight}
\ee
This is the negative binomial distribution $\mathrm{NB}(2\Delta,1-q)$, and the corresponding orthogonal polynomials are the Meixner polynomials, which satisfy
\begin{eqnarray}
&&
\sum_{x=0}^\infty
\frac{(\beta)_x}{x!}
c^x(1-c)^\beta
M_m(x;\beta,c)M_n(x;\beta,c)
=
\frac{n!c^{-n}}{(\beta)_n}\delta_{mn},
\\
&&
M_n(x;\beta,c)
=
\,_2F_1
\left(
-n,-x;\beta;1-\frac{1}{c}
\right),
\qquad
\beta>0,
\qquad
0<c<1 .
\end{eqnarray}
It follows that the orthonormal polynomials associated with \eqref{weight} are\footnote{Here $E=x+\Delta$.}
\be
p_n(x)
=
(-1)^n
\sqrt{
\frac{q^n(2\Delta)_n}{n!}
}
M_n(x;2\Delta,q).
\ee
From the Meixner recurrence relation
\be
(c-1)xM_n(x;\beta,c)
=
c(n+\beta)M_{n+1}(x;\beta,c)
-
(n+(n+\beta)c)M_n(x;\beta,c)
+
n M_{n-1}(x;\beta,c),
\ee
one reads off the Lanczos coefficients
\be
a_n
=
\frac{1+q}{1-q}(n+\Delta),
\qquad
b_{n+1}
=
\frac{\sqrt{q}}{1-q}
\sqrt{(n+1)(n+2\Delta)} .
\label{Lanczos}
\ee
The Krylov amplitudes are given by
\be
\varphi_n(t)
=
\langle K_n|\Psi(t)\rangle
=
\sum_x w_x\,
p_n(x)e^{-\im t(x+\Delta)} .
\ee
To evaluate this sum, we use the generating function of the Meixner polynomials,
\be
\sum_{x=0}^{\infty}
\frac{(\beta)_x}{x!}
y^x M_n(x;\beta,c)
=
(1-y)^{-\beta}
\left(
\frac{1-y/c}{1-y}
\right)^n .
\ee
Taking $\beta=2\Delta$, $c=q$, and $y=qe^{-\im t}$ gives
\be
\varphi_n(t)
=
e^{-\im \Delta t}
\sqrt{\frac{q^n(2\Delta)_n}{n!}}\,
\frac{
(1-q)^{2\Delta}
\left(1-e^{-\im t}\right)^n
}{
\left(1-q e^{-\im t}\right)^{n+2\Delta}
},
\ee
and 
\be
|\varphi_n(t)|^2
=
\frac{(2\Delta)_n}{n!}
\left[
\frac{
q|1-e^{-\im t}|^2
}{
|1-qe^{-\im t}|^2
}
\right]^n
\left[
\frac{
(1-q)^2
}{
|1-qe^{-\im t}|^2
}
\right]^{2\Delta}.
\ee
The probability distribution is again of negative-binomial form. To make this manifest, we define
\be
Q(t)
\equiv
\frac{
q|1-e^{-\im t}|^2
}{
|1-qe^{-\im t}|^2
}
=
\frac{
4q\sin^2(t/2)
}{
(1-q)^2+4q\sin^2(t/2)
}.
\label{qt}
\ee
In terms of $Q(t)$, the Krylov probability distribution becomes
\be
|\varphi_n(t)|^2
=
\frac{(2\Delta)_n}{n!}
\left[Q(t)\right]^n
\left[1-Q(t)\right]^{2\Delta}.
\ee
The spread complexity is therefore given by the mean of this distribution,
\be
C_{\text{bulk}}(t)
=
\sum_{n=0}^{\infty}
n|\varphi_n(t)|^2
=
\frac{2\Delta\, Q(t)}{1-Q(t)} .
\ee
Substituting \eqref{qt}, we find
\be
C_{\text{bulk}}(t)
=
\frac{8\Delta z_0^2}{(1-z_0^2)^2}
\sin^2\frac{t}{2},
\ee
in agreement with the boundary result \eqref{ckt}.

This agreement is not surprising. Through the extrapolation dictionary, the bulk and boundary calculations involve the same quantum state and the same Hamiltonian; the dictionary therefore acts here simply as a change of representation.
In the present local-quench setup, the main role of the CFT is to provide a particularly simple initial state. From the boundary perspective, this state is a generalized coherent state \cite{Perelomov:1986uhd,Zhang:1990fy}. This coherent-state structure will be crucial for the momentum--complexity relation discussed below. For a generic bulk state, no comparably simple analytic expression should be expected.
Once the initial state \eqref{bulkinitial} is specified, however, the extrapolation dictionary is no longer needed to carry out the bulk computation. The parameter $\rho_0$, related to $z_0$ by \eqref{z0rho}, admits the semiclassical interpretation of the initial position of the massive particle. It need not be large, so the particle probe need not stay close to the AdS boundary. More generally, holography is not required for spread complexity, or its growth rate, to admit a classical interpretation such as particle energy or momentum; this interpretation arises from the semiclassical limit of the quantum system itself.
This does not mean that any semiclassical wave packet will lead to the same behavior. One may choose a wave packet sharply peaked around some mean value, so that it admits a classical particle approximation, but the resulting spread complexity can still depend sensitively on the initial state. We will return to this point later with examples based on other simple symmetries.

Spread complexity is naturally suited to diagnosing quantum dynamics for a simple reason. Time evolution does not change the spectral statistics of the Hamiltonian. More generally, the statistics of any operator commuting with the Hamiltonian are time-independent. This is an immediate consequence of Ehrenfest's theorem for conserved quantities. Such spectral data therefore cannot detect the dynamical evolution of a state. One must instead consider an observable that does not commute with the Hamiltonian. Spread complexity is such an observable, and perhaps the simplest one constructed directly from the Hamiltonian. In the Krylov basis, the Hamiltonian is tridiagonal, reflecting the minimal noncommutativity built into the construction.

A similar question arises for entanglement entropy in the presence of nontrivial late-time bulk dynamics, such as the growth of the black hole interior \cite{Hartman:2013qma}. The saturation of entanglement entropy motivates the speculation that, in a suitable late-time regime, modular evolution becomes effectively aligned with physical time evolution. Under this assumption, nonconserved observables constructed from modular flow, such as modular Hamiltonian Krylov complexity \cite{Caputa:2023vyr} or the related Krylov entropy, could remain sensitive to late-time dynamics even after the entanglement entropy has saturated. It would be interesting to determine when such an alignment occurs and whether this mechanism admits a general holographic formulation.

\subsection{Classical limit}
We now explain how the classical description emerges from the bulk quantum description in the semiclassical limit.

The bulk state is a one-particle wave packet whose energy distribution is the negative binomial distribution $\mathrm{NB}(2\Delta,1-q)$. Its mean energy and variance are
\begin{eqnarray}
\langle H_{\text{bulk}}\rangle
&=&
\Delta+\langle n\rangle
=
\Delta\frac{1+q}{1-q}
=
\Delta\cosh\rho_0,
\label{expectE}
\\
\text{Var}(H_{\text{bulk}})
&=&
\langle n^2\rangle-\langle n\rangle^2
=
\frac{2\Delta q}{(1-q)^2}.
\label{varianceE}
\end{eqnarray}
The relative fluctuation is suppressed as $\Delta^{-1/2}$ at fixed $\rho_0$, so the state has a well-defined classical energy in the large-$\Delta$ limit. In this limit,
\be
m^2=\Delta(\Delta-1),
\qquad
\Delta\gg1
\quad\Rightarrow\quad
m\simeq\Delta .
\ee
Therefore,
\be
\langle H_{\text{bulk}}\rangle
\simeq
m\cosh\rho_0
\equiv
-p_t .
\ee
This is precisely the canonical energy of a classical massive particle released from rest at radial position $\rho_0$ in global AdS.

To see the trajectory emerge, consider the one-particle wavefunction
\be
\Psi(t,\rho)
=
{}_{\text{bulk}}\langle\Omega|
\hat{\phi}(t,\rho)
|\Psi_0\rangle
=
\sum_n c_n u_n ,
\label{oneparticle}
\ee
with $c_n$ and $u_n$ defined in \eqref{coe_cn} and \eqref{def_un}. The Gegenbauer generating function gives
\be
\Psi(t,\rho)
=
\mathcal{A}_\Delta
e^{-\im\Delta t}
\frac{
\sech^\Delta\rho
}{
\left(
1 - 2z_0 e^{-\im t} \tanh\rho + z_0^2 e^{-2\im t}
\right)^\Delta
},
\label{wavefun}
\ee
where
\be
\mathcal{A}_\Delta
=
\frac{
2^{\Delta-1} \Gamma(\Delta) (1-z_0^2)^\Delta
}{
\sqrt{\pi \Gamma(2\Delta)}
}.
\ee
In the large-$\Delta$ limit, $|\Psi(t,\rho)|^2$ is sharply localized at the saddle point
\be
\frac{d|\Psi(t,\rho)|^2}{d\rho}=0
\quad \Longrightarrow \quad
\tanh\rho_{\text{peak}}(t)
=
\frac{2z_0}{1+z_0^2}\cos t
=
\tanh\rho_0\cos t,
\ee
which is precisely the timelike geodesic \eqref{geodesic}. Thus the bulk state obtained from the extrapolation dictionary reduces, in the semiclassical limit, to a freely falling massive particle in global AdS. Although \eqref{oneparticle} is simply the bulk-to-bulk propagator, in the present context we interpret it as the projection of the quantum state onto the local-field representation. More generally, given any local operator $\hat{\mathcal{O}}(t,\rho)$, one may define a corresponding wave function by
\be 
\Psi_{\mathcal{O}}(t,\rho)
\equiv
{}_{\text{bulk}}\langle\Omega|
\hat{\mathcal{O}}(t,\rho)
|\Psi_0\rangle .
\ee

For a more general wave packet admitting a classical approximation, the wavefunction has a WKB form
\be
\Psi(t,\rho)
\sim
A(t,\rho)e^{\im S(t,\rho)} .
\ee
Substitution into the Klein-Gordon equation and expansion at leading order in the large-mass WKB approximation gives the Hamilton-Jacobi equation
\be
g^{\mu\nu}\partial_\mu S\partial_\nu S+m^2=0 .
\ee
For global AdS$_2$,
\be
-\frac{1}{\cosh^2\rho}
(\partial_t S)^2
+
(\partial_\rho S)^2
+
m^2
=
0 .
\ee
With
\be
p_t=\partial_t S=-E,
\qquad
p_\rho=\partial_\rho S,
\ee
this becomes
\be
-\frac{E^2}{\cosh^2\rho}
+
p_\rho^2
+
m^2
=
0 .
\ee
The effective particle Hamiltonian is therefore
\be
H_p
=
E
=
\cosh\rho
\sqrt{p_\rho^2+m^2}.
\ee
Hamilton's equations are
\begin{eqnarray}
\dot{\rho}
&=&
\frac{\partial H_p}{\partial p_\rho}
=
\frac{\cosh\rho\,p_\rho}{\sqrt{p_\rho^2+m^2}},
\\
\dot{p}_\rho
&=&
-\frac{\partial H_p}{\partial \rho}
=
-\sinh\rho
\sqrt{p_\rho^2+m^2}.
\end{eqnarray}
Taking one more time derivative gives
\be
\ddot{p}_\rho+p_\rho=0 .
\label{eom_phase}
\ee
Thus the classical trajectory emerges by projecting $|\Psi(t)\rangle$ onto the representation associated with the field operator $\hat{\phi}$.

The same quantum state can alternatively be described in the spread-complexity representation, defined by the eigenbasis of the operator $\hat C$ introduced below. In the local-field representation, this state becomes localized in the semiclassical limit.
The same holds in the complexity representation. Indeed, $|\varphi_n(t)|^2$ is again a negative binomial distribution, $\text{NB}(2\Delta,1-Q(t))$, and its relative fluctuation vanishes as $\Delta\to\infty$. This suggests that the wavefunction
\be
\Psi^C(t)
\equiv
{}_{\text{bulk}}\langle \Omega|\hat{C}|\Psi(t)\rangle,
\qquad
\hat{C}
=
\sum_n n|K_n\rangle\langle K_n|,
\ee
admits a WKB approximation. In this limit the operator $\hat C$ reduces to a classical variable $C(p_\rho,\rho)$, namely a function on the phase space of the free particle in AdS$_2$. In the next subsection we obtain this function directly by quantizing a free particle in AdS$_2$.

The equation of motion for spread complexity can also be derived from the algebraic structure of the Krylov chain. The Lanczos coefficients \eqref{Lanczos} imply that, in the Krylov basis,
\be
H
=
\alpha \hat{\mathcal{L}}_0
+
\frac{\beta}{2}
\left(
\hat{\mathcal{L}}_+
+
\hat{\mathcal{L}}_-
\right),
\qquad
\alpha
=
\frac{1+q}{1-q},
\qquad
\beta
=
\frac{2\sqrt q}{1-q}.
\label{ham}
\ee
Here $\{\hat{\mathcal{L}}_0,\hat{\mathcal{L}}_\pm\}$ generate $SL(2,\mathbb{R})$ and satisfy
\begin{eqnarray}
&&
[\hat{\mathcal{L}}_0,\hat{\mathcal{L}}_\pm]
=
\mp \hat{\mathcal{L}}_\pm,
\qquad
[\hat{\mathcal{L}}_+,\hat{\mathcal{L}}_-]
=
2\hat{\mathcal{L}}_0,
\label{sl2rcm}
\\
&&
\hat{\mathcal{L}}_0|K_n\rangle
=
(n+\Delta)|K_n\rangle,
\\
&&
\hat{\mathcal{L}}_-|K_n\rangle
=
\sqrt{(n+1)(n+2\Delta)}|K_{n+1}\rangle,
\\
&&
\hat{\mathcal{L}}_+|K_n\rangle
=
\sqrt{n(n+2\Delta-1)}|K_{n-1}\rangle .
\end{eqnarray}
The complexity operator is
\be
\hat C
=
\hat{\mathcal{L}}_0-\Delta .
\ee
A related $SL(2,\mathbb{R})$ construction for operator growth in two-dimensional CFTs appears in \cite{Caputa:2021sib}.
Using the Heisenberg equation
\be
\dot{\hat C}
=
\im[H,\hat C]
\ee
twice gives
\be
\ddot{\hat C}
=
\alpha H-\hat{\mathcal{L}}_0,
\ee
or
\be
\ddot{\hat C}+\hat C
=
\alpha H-\Delta .
\ee
Taking the expectation value in $|K_0\rangle$,
\begin{eqnarray}
\ddot C+C
&=&
\alpha\langle H\rangle-\Delta
=
(\alpha^2-1)\Delta,
\\
\Rightarrow\qquad
C(t)
&=&
(\alpha^2-1)\Delta(1-\cos t).
\label{classicalC}
\end{eqnarray}
Thus spread complexity behaves as the shifted position of a harmonic oscillator. Comparing this equation with \eqref{eom_phase}, we see that the classical dynamics of spread complexity matches that of the free particle.

The preceding derivation also indicates that the classical spread complexity $C(\rho,p_\rho)$ is an example of a complexity observable in the CAny proposal \cite{Belin:2021bga,Belin:2022xmt}. There are, however, two differences: CAny is usually formulated in eternal black hole spacetimes and in terms of observables of the gravitational theory, whereas here we consider an observable on the phase space of a probe in global AdS.

The proposed analogy is as follows. In CAny, the scalar function $F_2$\footnote{For the definitions of $F_2$ and $\co_{F_1,\Sigma_{F_2}}$, see \cite{Belin:2021bga,Belin:2022xmt}.} specifies geometric data entering the observable $\co_{F_1,\Sigma_{F_2}}$. In the present setting, the corresponding classical input is the phase space of a probe, such as a particle, string, or brane, and the observable is a function on that phase space. For the observable to diagnose nontrivial dynamics, it should not commute with the Hamiltonian. Spread complexity provides a simple example of this kind.

Once $\co_{F_1,\Sigma_{F_2}}$ is identified as a phase-space function, one can try to quantize it, for instance by geometric quantization \cite{10.1093/oso/9780198536734.001.0001}. In favorable cases, as for spread complexity, the quantized observable becomes a genuine quantum complexity measure. The qualification is important: not every classical observable admits a consistent quantization. Thus not every observable appearing in the CAny proposal is necessarily well defined at the quantum level, because of the Groenewold-van Hove no-go theorem \cite{groenewold12principles,van1951certaines}. We leave a fuller discussion of this issue to future work.

One may also consider a non-free particle probe. Equivalently, one may evolve with functions other than the global Hamiltonian. The spread complexity in \eqref{classicalC} is periodic because the Hamiltonian \eqref{ham} belongs to the elliptic class. Instead consider
\be
H_b
=
-\frac{\im}{2}
\left(
\hat{\mathcal{L}}_+
-
\hat{\mathcal{L}}_-
\right).
\ee
For simplicity, set $\rho_0=0$, so that the initial state is the primary state of $\hat{\mathcal{L}}_0$:
\be
|\psi_0\rangle
=
\co(0)|\Omega\rangle
=
|\Delta\rangle .
\ee
The spread complexity of this system has been studied, for example, in \cite{Balasubramanian:2022tpr}. One finds
\begin{eqnarray}
|\psi(t)\rangle
&=&
e^{-\frac{1}{2}t(\hat{\mathcal{L}}_+-\hat{\mathcal{L}}_-)}
|\Delta\rangle
\nonumber\\
&=&
\frac{1}{\cosh^{2\Delta}(t/2)}
\sum_{n=0}^\infty
\tanh^n\frac{t}{2}
\sqrt{
\frac{\Gamma(2\Delta+n)}{n!\Gamma(2\Delta)}
}
|\Delta,n\rangle,
\\
C(t)
&=&
\Delta(\cosh t-1).
\label{bC}
\end{eqnarray}
Here
\be
|\varphi_n|^2
=
\frac{(2\Delta)_n}{n!}
\tanh^{2n}\frac{t}{2}
\left(
1-\tanh^2\frac{t}{2}
\right)^{2\Delta},
\ee
which is again $\text{NB}(2\Delta,1-\tanh^2(t/2))$. Hence, the distribution becomes sharply peaked as $\Delta\to\infty$ and admits a classical approximation. The Heisenberg equation gives
\be
\ddot C
=
C+\Delta,
\ee
whose solution is \eqref{bC}.

From the boundary point of view, the evolution moves the operator insertion from $z_0$ to $\tilde z$:
\be
e^{-\frac{1}{2}t(\hat{\mathcal{L}}_+-\hat{\mathcal{L}}_-)}
\co(z_0)
e^{\frac{1}{2}t(\hat{\mathcal{L}}_+-\hat{\mathcal{L}}_-)}
\sim
\co(\tilde z),
\qquad
\tilde z
=
\frac{
z_0+\tanh(t/2)
}{
1+z_0\tanh(t/2)
}
\Bigg|_{z_0=0}
=
\tanh\frac{t}{2}.
\ee
The one-particle wavefunction is therefore
\be
\Psi(t,\rho)
\sim
\frac{
\sech^\Delta\rho
}{
\left(
1-2\tilde z\tanh\rho+\tilde z^2
\right)^\Delta
},
\ee
which is localized at
\be
\tanh\rho_{\rm peak}(t)
=
\frac{2\tilde z}{1+\tilde z^2}
=
\tanh t,
\qquad
\Rightarrow
\qquad
\rho_{\rm peak}(t)=t .
\ee
Thus $H_b$ pushes the particle away from the center along a radial boost orbit. Since the particle is localized at $\rho_{\rm peak}(t)$, its global AdS energy is
\be
E_{\rm global}(t)
=
\Delta\cosh\rho_{\rm peak}(t).
\ee
Under this evolution, spread complexity measures the global energy above the rest energy:
\be
C(t)
=
\Delta(\cosh t-1)
=
E_{\rm global}(t)-\Delta .
\ee

\subsection{Quantization of a free massive particle in \texorpdfstring{AdS$_2$}{AdS2}}
Instead of deriving the particle representation as the semiclassical limit of a scalar field theory, we now quantize a free massive particle in AdS$_2$ directly, following \cite{ElGradechi:1992te}. This serves two purposes. First, it gives an explicit example of quantizing a complexity observable in the CAny proposal: the particle provides the phase space, and the observable corresponding to spread complexity is obtained by quantizing a classical function on this phase space. Second, the construction uses only the bulk theory and does not assume AdS/CFT. It can therefore be generalized, at least in principle, to spacetimes without a known field-theory dual.

Embed AdS$_2$ in $\mathbb{R}^{2,1}$ with metric
$\eta_{\mu\nu}=\operatorname{diag}(-1,-1,1)$ and coordinates $y^\mu$ satisfying
\be
-(y^0)^2-(y^1)^2+(y^2)^2=-1 .
\ee
The action of a relativistic massive particle is
\be
S
=
\int d\tau
\left[
\frac{\mathbf m}{2}
\eta_{\mu\nu}\dot y^\mu\dot y^\nu
-
\frac{\Lambda}{2}
\left(
\eta_{\mu\nu}y^\mu y^\nu+1
\right)
\right],
\ee
where $\dot y^\mu\equiv dy^\mu/d\tau$ and $\Lambda$ is a Lagrange multiplier. Varying with respect to $y^\mu$ gives
\be
\frac{d^2y^\mu}{d\tau^2}+y^\mu=0,
\ee
where $\Lambda=\mathbf m$ has been used, as follows from
\be
\eta_{\mu\nu}\dot y^\mu\dot y^\nu=-1 .
\ee
The general solution is
\be
y^\mu(\tau)
=
e_1^\mu\cos\tau+e_2^\mu\sin\tau,
\label{solution}
\ee
with
\be
\eta_{\mu\nu}e_1^\mu e_1^\nu
=
\eta_{\mu\nu}e_2^\mu e_2^\nu
=
-1,
\qquad
\eta_{\mu\nu}e_1^\mu e_2^\nu=0.
\ee
The physical phase space is two-dimensional, so the parametrization contains one redundancy. It comes from the shift $\tau\to\tau+\tau_0$, which induces
\be
(e_1^\mu,e_2^\mu)
\to
\left(
\cos\tau_0\,e_1^\mu+\sin\tau_0\,e_2^\mu,
\,
\cos\tau_0\,e_2^\mu-\sin\tau_0\,e_1^\mu
\right).
\ee
In \cite{ElGradechi:1992te}, this gauge freedom is fixed by imposing $y^0=0$, using the fact that each solution intersects the surface $y^0=0$ exactly once. We choose a different gauge, better adapted to coordinates with direct physical meaning.

In global AdS$_2$ coordinates $(t,\rho)$,
\be
ds^2=-\cosh^2\rho\,dt^2+d\rho^2,
\ee
the embedding is
\be
y^0
=
\cosh\rho\sin t,
\qquad
y^1
=
\cosh\rho\cos t,
\qquad
y^2
=
\sinh\rho .
\label{embed}
\ee
The geodesic in global coordinates is
\be
\tanh\rho
=
\tanh\rho_0\cos(t-t_0),
\label{geodesic_particle}
\ee
so $(t_0,\rho_0)$ are natural coordinates on phase space. Substituting \eqref{solution} into \eqref{embed},
\begin{eqnarray}
\sinh\rho(\tau)
&=&
e_1^2\cos\tau+e_2^2\sin\tau,
\\
\tan t(\tau)
&=&
\frac{
e_1^0\cos\tau+e_2^0\sin\tau
}{
e_1^1\cos\tau+e_2^1\sin\tau
}.
\end{eqnarray}
Comparison with \eqref{geodesic_particle} gives
\begin{eqnarray}
\sinh\rho_0
&=&
\sqrt{(e_1^2)^2+(e_2^2)^2},
\\
\tan t_0
&=&
\frac{
e_2^2 e_1^1-e_1^2 e_2^1
}{
e_1^2 e_2^0-e_2^2 e_1^0
}.
\end{eqnarray}
The presymplectic form is
\begin{eqnarray}
\omega
&=&
dp_\mu\wedge dy^\mu
=
\mathbf m\,\eta_{\mu\nu}d\dot y^\mu\wedge dy^\nu
\nonumber\\
&=&
\mathbf m
\left(
de_1^0\wedge de_2^0
+
de_1^1\wedge de_2^1
-
de_1^2\wedge de_2^2
\right),
\end{eqnarray}
which is independent of time. Using the $\tau$-translation symmetry, we choose the gauge such that $t=t_0$ at $\tau=0$. Then
\be
e_1^\mu
=
y^\mu(0)
=
\left(
\cosh\rho_0\sin t_0,
\cosh\rho_0\cos t_0,
\sinh\rho_0
\right).
\ee
Solving the constraints gives\footnote{We have discarded the other branch, $e_2=(\cos t_0,-\sin t_0,0)$.}
\be
e_2^\mu
=
\left(
-\cos t_0,
\sin t_0,
0
\right).
\ee
The symplectic form in $(\rho_0,t_0)$ coordinates is
\be
\omega
=
\mathbf m\sinh\rho_0\,d\rho_0\wedge dt_0 .
\label{sympar}
\ee
The conserved charges
$L_{\mu\nu}=\mathbf m(y_\mu\dot y_\nu-y_\nu\dot y_\mu)$ associated with the $SO(2,1)$ isometries are
\be
L_{01}
=
\mathbf m\cosh\rho_0,
\qquad
L_{02}
=
-\mathbf m\sinh\rho_0\cos t_0,
\qquad
L_{12}
=
\mathbf m\sinh\rho_0\sin t_0 .
\ee
The global energy is
\be
E=-p_t=\mathbf m\cosh\rho_0=L_{01}.
\ee
The parameters $(t_0,\rho_0)$ can be combined into a complex coordinate on
the unit disk,
\be
z
=
\tanh\frac{\rho_0}{2}\,
e^{\im t_0},
\qquad |z|<1 .
\ee
In terms of this coordinate, the classical phase space is identified with the Poincar\'e disk, and the Kirillov--Kostant--Souriau (KKS) symplectic form is the standard $SL(2,\mathbb{R})$-invariant K\"ahler form:
\be
\omega
=
\mathbf m\,
\frac{
2\im\,dz\wedge d\bar z
}{
(1-|z|^2)^2
}.
\label{sl2r}
\ee
Geometric quantization gives a Hilbert space of holomorphic functions with an inner product (assuming $\mathbf m>1/2$)
\be
\langle f,g\rangle
=
\frac{2\mathbf m-1}{\pi}
\int_{|z|\leq 1}
\overline{f(z)}g(z)
(1-|z|^2)^{2\mathbf m-2}\,d^2z .
\ee
An orthonormal basis is
\be
\psi_n
=
\sqrt{
\frac{(2\mathbf m)_n}{n!}
}
z^n,
\qquad
n=0,1,\dots .
\ee
The conserved charges are quantized as
\begin{eqnarray}
L_{01}
&\to&
\hat l_0
=
z\partial_z+\mathbf m,
\\
-L_{02}+\im L_{12}
&\to&
\hat l_-
=
z^2\partial_z+2\mathbf m z,
\\
-L_{02}-\im L_{12}
&\to&
\hat l_+
=
\partial_z .
\end{eqnarray}
These operators obey the same commutation relations as \eqref{sl2rcm}. We distinguish $\hat{\mathcal L}_n$ from $\hat l_n$: the former act on the Krylov subspace, while the latter are associated with spacetime isometries. In particular, $\hat l_0=H$ generates global time translations. The two sets are related by an $SL(2,\mathbb R)$ conjugation.
Let $|n\rangle$ denote the states whose wave functions in the holomorphic
disk representation are $\psi_n(z)$,
\be
\langle z|n\rangle
\equiv
\psi_n(z) .
\ee
The differential-operator action of the generators on
$\psi_n(z)$ induces the following action on the abstract states:
\begin{eqnarray}
\langle z|\hat l_k|n\rangle
&=&
\hat l_k\psi_n(z),
\qquad
\hat l_0|n\rangle
=
(\mathbf m+n)|n\rangle,
\\
\hat l_-|n\rangle
&=&
\sqrt{(2\mathbf m+n)(n+1)}\,|n+1\rangle,
\\
\hat l_+|n\rangle
&=&
\sqrt{n(2\mathbf m+n-1)}\,|n-1\rangle .
\end{eqnarray}
Thus, the quantization of a free particle in AdS$_2$ gives a discrete-series representation of $SL(2,\mathbb{R})$, which maps to the one-particle sector of the massive scalar field discussed above.

To obtain a state admitting a clear classical interpretation, it is natural to choose the generalized coherent state \cite{Perelomov:1986uhd,Zhang:1990fy}
\be
|\psi(u)\rangle
=
(1-|u|^2)^{\mathbf m}
\sum_{n=0}^{\infty}
\sqrt{
\frac{(2\mathbf m)_n}{n!}
}
u^n|n\rangle .
\ee
Its holomorphic wavefunction is
\begin{eqnarray}
\langle z|\psi(u)\rangle
&=&
(1-|u|^2)^{\mathbf m}
\sum_{n=0}^{\infty}
\frac{(2\mathbf m)_n}{n!}
u^n z^n
\nonumber\\
&=&
\frac{
(1-|u|^2)^{\mathbf m}
}{
(1-zu)^{2\mathbf m}
}.
\end{eqnarray}
For this state, the probability distribution in the $\hat l_0=H$ eigenbasis is
\begin{eqnarray}
\langle \psi(u)|\hat l_0|\psi(u)\rangle
&=&
\sum_n
(\mathbf m+n)
(1-|u|^2)^{2\mathbf m}
|u|^{2n}
\frac{(2\mathbf m)_n}{n!}
\nonumber\\
&\equiv&
\sum_n(\mathbf m+n)w_n,
\end{eqnarray}
where $w_n$ is $\text{NB}(2\mathbf m,1-|u|^2)$. This agrees with \eqref{weight} after identifying
\be
|u|=|z_0|=\tanh\frac{\rho_0}{2},
\qquad
\Delta=\mathbf{m} .
\ee
The previous results are therefore reproduced. The one-particle state $a_n^\dagger|\Omega\rangle_{\text{bulk}}$ can be identified with the Fock state $|n\rangle$. Since $\mathcal{O}|\Omega\rangle$ is expanded in one-particle states by the extrapolation dictionary, it may also be expanded in the Fock basis.

Therefore, the time-evolved coherent state
\be
|\psi(t,u)\rangle
\equiv
e^{-\im\hat l_0 t}|\psi(u)\rangle
\ee
is dual to the geodesic \eqref{geodesic}. We have already identified $|u|=\tanh(\rho_0/2)$. To fix the phase, use the fact that, for a coherent state, the expectation values of the quantum generators agree with the corresponding classical charges. Matching
\be
-L_{02}\pm\im L_{12}
=
\mathbf m\sinh\rho_0 e^{\pm\im t_0}
=
\langle \psi(u)|\hat l_{\mp}|\psi(u)\rangle,
\ee
we find
\be
u
=
\bar z
=
\tanh\frac{\rho_0}{2}
e^{-\im t_0}.
\label{uzbar}
\ee
Thus, the boundary state $\co(z)|\Omega\rangle$ is exactly dual to the generalized coherent state $|\psi(u)\rangle$ with \eqref{uzbar}. The classical observables are angular momenta, equivalently the conserved charges associated with Killing vectors. These charges are equal to the expectation values of the corresponding quantum operators.

The spread complexity operator is $\hat{\mathcal L}_0-\Delta$:
\be
C(t)
=
\langle \psi(t,u)|
\hat{\mathcal L}_0
|\psi(t,u)\rangle
-\Delta .
\ee
From \eqref{ham}, the physical Hamiltonian can be written as
\be
\hat l_0
=
H
=
\cosh\rho_0\,\hat{\mathcal L}_0
+
\frac{\sinh\rho_0}{2}
\left(
\hat{\mathcal L}_+
+
\hat{\mathcal L}_-
\right),
\ee
where $|u|=\tanh(\rho_0/2)$ has been used. Inverting gives
\be
\hat{\mathcal L}_0
=
\cosh\rho_0\,\hat l_0
-
\frac{\sinh\rho_0}{2}
\left(
e^{\im t_0}\hat l_+
+
e^{-\im t_0}\hat l_-
\right),
\label{Qspread}
\ee
which is consistent with
\begin{eqnarray}
\langle \psi(u)|\hat{\mathcal L}_0|\psi(u)\rangle
&=&
\left\langle \psi(u)\left|
\cosh\rho_0\,\hat l_0
-
\frac{\sinh\rho_0}{2}
\left(
e^{\im t_0}\hat l_+
+
e^{-\im t_0}\hat l_-
\right)
\right|\psi(u)\right\rangle
\nonumber\\
&=&
\cosh\rho_0\,L_{01}
+
\sinh\rho_0
\left(
\cos t_0\,L_{02}
-
\sin t_0\,L_{12}
\right)
\nonumber\\
&=&
\Delta .
\end{eqnarray}
Therefore, spread complexity is a linear combination of the particle angular momenta:
\be
C(t)
=
\cosh\rho_0\,L_{01}
+
\sinh\rho_0
\left[
\cos(t-t_0)L_{02}
+
\sin(t-t_0)L_{12}
\right]
-\Delta .
\ee
The time dependence follows from the fact that the evolution generated by $\hat l_0$ rotates the ladder generators by a phase:
\be
e^{\im t\hat l_0}\hat l_\pm e^{-\im t\hat l_0}
=
e^{\mp\im t}\hat l_\pm .
\ee
Therefore, the phases appearing in \eqref{Qspread} acquire the shift
\be
t_0\rightarrow t_0-t .
\ee

The preceding construction gives a direct phase-space interpretation of the spread complexity operator and of the momentum--complexity proposal of \cite{Caputa:2024sux,Li:2025fqz}. It also suggests a probe analogue of a CAny observable. In CAny, complexity observables are constructed from geometric data, such as lengths associated with geodesics. In the present case, the underlying classical space is instead the phase space of a free-particle probe, parametrized by $(t_0,\rho_0)$ and equipped with the symplectic form \eqref{sympar}. The corresponding probe observable is a particular linear combination of the $SL(2,\mathbb{R})$ angular momenta. Upon quantization, and when evaluated in a coherent initial state, it becomes the spread complexity operator \eqref{Qspread}; its expectation value gives the spread complexity.

We close with a comment on the physical origin of spread complexity. For a generic initial wave packet, several effects can contribute to its growth. In a closed quantum system, the two basic contributions are coherent motion of the packet and dispersive spreading. In an open system, particle creation and loss may also contribute. In the example studied here, the probe phase space is a coadjoint orbit of a Lie group, the initial state is the corresponding generalized coherent state, and the Hamiltonian lies in the same Lie algebra. The packet therefore preserves its shape and does not disperse. Spread complexity is then governed entirely by coherent motion. This is the reason it can be identified with classical particle dynamics. Equivalently, the variance of the probability distribution associated with the one-particle wavefunction \eqref{wavefun} vanishes in the large-$\Delta$ limit. In later examples, dispersive spreading need not vanish in the classical limit and can instead dominate the spread complexity.

\section{Spread complexity in flat spacetime}
\label{sec_flat}
As a simple example with analytic control, consider a free massive particle of mass $M$ in two-dimensional flat spacetime,
\be
ds^2=-dt^2+dx^2 .
\ee
The geodesic is
\be
t=-\frac{p_t}{M}\tau+t_0,
\qquad
x=\frac{p_x}{M}\tau+x_0 .
\ee
We choose the gauge in which $\tau=0$ corresponds to $t=0$. Then $t_0=0$, and the reduced phase space is parametrized by
\be
p_x\equiv p,
\qquad
x_0\equiv q,
\ee
with symplectic form
\be
\omega=dp\wedge dq .
\ee
Canonical quantization gives
\be
\mathcal{H}_{\text{flat}}
=
L^2(\mathbb{R},dp)
=
L^2(\mathbb{R},dq).
\ee
We introduce generalized eigenstates $|q\rangle$ satisfying
\be
\hat q|q\rangle=q|q\rangle,
\qquad
\langle q'|q\rangle=\delta(q-q'),
\qquad
\int_{\mathbb R}dq\,|q\rangle\langle q|=\mathbf 1 .
\ee
A general state is expanded as
\be
|f\rangle
=
\int_{\mathbb R}dq\,f(q)|q\rangle,
\qquad
f(q)=\langle q|f\rangle .
\ee
With $[\hat q,\hat p]=\im$, translations are generated by $\hat p$:
\be
|q\rangle=e^{-\im \hat p q}|0\rangle .
\ee
Indeed,
\be
\hat q e^{-\im \hat p q}|0\rangle
=
e^{-\im \hat p q}(\hat q+q)|0\rangle
=
q|q\rangle .
\ee
Differentiating the translated state gives the action of momentum on the position kets,
\be
\hat p|q\rangle=\im\partial_q|q\rangle,
\ee
or equivalently, on wavefunctions,
\be
\langle q|\hat p|f\rangle=-\im\partial_q f(q).
\ee
It is also useful to introduce the dimensionless complex coordinates
\be
z=\frac{p/M+\im q M}{\sqrt{2}},
\qquad
\bar z=\frac{p/M-\im q M}{\sqrt{2}},
\ee
where $M$ is an arbitrary reference scale. For $\omega=dp\wedge dq$, one finds
\be
\omega=\im dz\wedge d\bar z .
\ee
K\"ahler quantization in the holomorphic polarization gives the Fock Hilbert space, with basis states represented by
\be
\psi_n(z)
=
\langle z|n\rangle
=
\frac{z^n}{\sqrt{n!}},
\qquad n=0,1,2,\cdots .
\ee
The elementary operators act as
\begin{eqnarray}
\hat z\psi_n(z)
&=&
z\psi_n(z)
=
\langle z|a^\dagger|n\rangle ,
\nonumber\\
\hat{\bar z}\psi_n(z)
&=&
\partial_z\psi_n(z)
=
\langle z|a|n\rangle .
\end{eqnarray}

\subsection{Coherent state and Gaussian packet}
In contrast to the AdS case, this example will show that when the physical Hamiltonian is not an element of the symmetry algebra preserving the coherent-state manifold, the spread complexity of a coherent initial state probes the dispersive dynamics of the wave packet rather than a classical momentum.

Take the initial state to be the coherent state $|0\rangle$, whose wavefunction in the $q$-representation is the Gaussian packet
\be
\langle q|0\rangle
=
\left(\frac{M}{\sqrt{\pi}}\right)^{1/2}
\exp\left[-\frac{M^2q^2}{2}\right],
\ee
with
\be
\hat p
=
\frac{M}{\sqrt{2}}(a+a^\dagger),
\qquad
\hat q
=
\frac{\im}{\sqrt{2}M}(a-a^\dagger),
\qquad
[\hat q,\hat p]=\im .
\ee
Motivated by the AdS case, we first choose a toy Hamiltonian belonging to the Heisenberg algebra,
\be
H_{\rm toy}
=
a+a^\dagger
=
\sqrt{2}\frac{\hat p}{M}.
\label{artiH}
\ee
It generates translations in position space:
\be
e^{-\im H_{\rm toy}\tau}|q_1\rangle
=
\left|q_1+\sqrt{2}\frac{\tau}{M}\right\rangle .
\ee
The time-evolved wavefunction is therefore
\begin{eqnarray}
\langle q|\psi(\tau)\rangle
&=&
\left\langle q-\sqrt{2}\frac{\tau}{M}\middle|0\right\rangle
\nonumber\\
&=&
\left(\frac{M}{\sqrt{\pi}}\right)^{1/2}
\exp\left[
-\frac{M^2}{2}
\left(q-\sqrt{2}\frac{\tau}{M}\right)^2
\right],
\end{eqnarray}
while in momentum space
\be
\langle p|\psi(\tau)\rangle
=
\left(\frac{1}{M\sqrt{\pi}}\right)^{1/2}
\exp\left[
-\frac{p^2}{2M^2}
-\im\sqrt{2}\frac{p\tau}{M}
\right].
\ee
Thus $H_{\rm toy}$ translates the packet center without producing dispersion.

The Krylov basis generated by $H_{\rm toy}$ can be identified with the Fock basis. Indeed, since
\be
H_{\rm toy}=a+a^\dagger
\ee
generates a coherent displacement, one has
\be
|\psi(\tau)\rangle
=
e^{-\im H_{\rm toy}\tau}|0\rangle
=
e^{-\im \tau(a+a^\dagger)}|0\rangle
=
e^{-\tau^2/2}
\sum_{n=0}^{\infty}
\frac{(-\im \tau)^n}{\sqrt{n!}}|n\rangle .
\ee
Thus the Krylov amplitudes are
\be
\varphi_n(\tau)
=
e^{-\tau^2/2}
\frac{(-\im \tau)^n}{\sqrt{n!}},
\ee
and the Krylov probabilities form a Poisson distribution,
\be
|\varphi_n(\tau)|^2
=
e^{-\tau^2}\frac{\tau^{2n}}{n!}.
\ee
The corresponding spread complexity and its growth rate are
\be
C(\tau)
=
\sum_{n=0}^{\infty}n|\varphi_n(\tau)|^2
=
\tau^2,
\qquad
\frac{dC}{d\tau}
=
2\tau .
\label{flatC}
\ee
The same result follows at the operator level. The spread complexity operator is
\be
\hat C=a^\dagger a .
\ee
Its evolution obeys
\begin{eqnarray}
\frac{d\hat C}{d\tau}
&=&
\im[H_{\rm toy},a^\dagger a]
=
\sqrt{2}M\hat q,
\qquad
\frac{d^2\hat C}{d\tau^2}=2 .
\end{eqnarray}
In this case the growth rate of spread complexity is proportional to the position operator rather than the momentum operator. This is natural. Under $H_{\rm toy}$, the momentum, and hence the physical energy in flat space, is conserved. The observable that detects the nontrivial dynamics is the position of the packet center.

We now connect this quantum-mechanical discussion to the one-particle sector of a relativistic scalar field in flat spacetime. The Klein-Gordon field has the mode expansion
\be
\hat\phi(t,x)
=
\int_{-\infty}^{\infty}
\frac{dp}{\sqrt{2E_p}}
\left[
\hat a_p e^{-\im E_p t+\im px}
+
\hat a_p^\dagger e^{\im E_p t-\im px}
\right],
\ee
where
\be
E_p=\sqrt{p^2+M^2}.
\ee
For the toy-evolved coherent packet, the corresponding one-particle wavefunction is
\begin{eqnarray}
\Psi(\tau,t,x)
&\equiv&
{}_{\rm bulk}\langle \Omega|
\hat\phi(t,x)
|\psi(\tau)\rangle
=
\int_{-\infty}^{\infty}
\frac{dp}{\sqrt{2E_p}}
e^{\im px-\im E_p t}
\langle p|\psi(\tau)\rangle
\nonumber\\
&=&
\frac{1}{\sqrt{M}\pi^{1/4}}
\int_{-\infty}^{\infty}
\frac{dp}{\sqrt{2E_p}}
\exp\left[
-\frac{p^2}{2M^2}
+\im p\left(x-\sqrt{2}\frac{\tau}{M}\right)
-\im E_p t
\right].
\label{exact_wavepacket}
\end{eqnarray}
Here $t$ is the physical time generated by the relativistic Hamiltonian, whereas $\tau$ is the parameter generated by the toy Hamiltonian \eqref{artiH}. In the nonrelativistic regime $|p|\ll M$,
\be
E_p\simeq M+\frac{p^2}{2M}.
\ee
Up to normalization, the Gaussian integral gives
\begin{eqnarray}
\Psi_{\rm NR}(\tau,t,x)
&\simeq&
\mathcal{N}(t)
e^{-\im Mt}
\exp\left[
-\frac{\left(x-\sqrt{2}\tau/M\right)^2}
{2\left(1/M^2+\im t/M\right)}
\right],
\end{eqnarray}
where $\mathcal{N}(t)$ is fixed by normalization. The normalized probability profile is
\be
|\Psi_{\rm NR}(\tau,t,x)|^2
=
\frac{1}{\sqrt{\pi}\sigma(t)}
\exp\left[
-\frac{\left(x-x_c(\tau)\right)^2}{\sigma(t)^2}
\right],
\label{prob_density}
\ee
with
\be
x_c(\tau)=\sqrt{2}\frac{\tau}{M},
\qquad
\sigma(t)=\sqrt{\frac{1}{M^2}+t^2}.
\label{wavewidth}
\ee
Thus the toy Hamiltonian controls the motion of the center, while the physical Hamiltonian controls the dispersion of the packet.

Now consider the physical Hamiltonian itself,
\be
H=E_p=\sqrt{p^2+M^2}.
\ee
We compute the associated spread complexity in the nonrelativistic regime, where the initial Gaussian momentum distribution is supported predominantly on $|p|\ll M$. In this approximation,
\be
H\simeq M+\epsilon,
\qquad
\epsilon=\frac{p^2}{2M}.
\ee
As before, we first determine the spectral distribution and then construct the orthogonal polynomials defining the Krylov basis. The initial momentum distribution is
\be
\rho(p)dp
=
|\langle p|0\rangle|^2dp
=
\frac{1}{M\sqrt{\pi}}
e^{-p^2/M^2}dp .
\ee
The spectral variable within this approximation is therefore the kinetic energy $\epsilon$.
Taking into account the two branches
\be
p=\pm\sqrt{2M\epsilon},
\ee
we obtain
\begin{eqnarray}
\rho(\epsilon)d\epsilon
&=&
\frac{1}{M\sqrt{\pi}}
e^{-2\epsilon/M}
\cdot
2\frac{dp}{d\epsilon}d\epsilon
\nonumber\\
&=&
\frac{\sqrt{2}}{\sqrt{\pi M}}
\epsilon^{-1/2}
e^{-2\epsilon/M}d\epsilon .
\end{eqnarray}
This is a gamma distribution,
\be
\rho(\epsilon)d\epsilon
=
\frac{1}{\Gamma(k)\theta^k}
\epsilon^{k-1}e^{-\epsilon/\theta}d\epsilon,
\ee
with
\be
k=\frac{1}{2},
\qquad
\theta=\frac{M}{2}.
\label{gamma_dist}
\ee

The gamma distribution is the orthogonality measure for the generalized Laguerre polynomials:
\begin{eqnarray}
\int_0^\infty
\rho(\epsilon)P_n(\epsilon)P_m(\epsilon)d\epsilon
=
\delta_{nm},
\qquad
P_n(\epsilon)
\equiv
(-1)^n
\sqrt{\frac{n!\Gamma(k)}{\Gamma(n+k)}}
L_n^{(k-1)}
\left(\frac{\epsilon}{\theta}\right).
\end{eqnarray}
The Krylov basis can be written as
\be
K_n(\epsilon)
\equiv
\langle \epsilon|K_n\rangle
=
\sqrt{\rho(\epsilon)}P_n(\epsilon).
\ee
The three-term recurrence relation
\be
L_{n+1}^{(k-1)}(x)
=
\frac{(2n+k-x)L_{n}^{(k-1)}(x)-(n+k-1)L_{n-1}^{(k-1)}(x)}
{n+1}
\ee
gives the Lanczos coefficients
\be
a_n
=
\theta(2n+k)
=
Mn+\frac{M}{4},
\qquad
b_n
=
\theta\sqrt{n(n+k-1)}
=
\frac{M}{2}
\sqrt{n\left(n-\frac{1}{2}\right)} .
\ee
The rest-mass term in the full relativistic Hamiltonian only shifts
\be
a_n\rightarrow M+a_n .
\ee
It therefore contributes only an overall phase and does not affect the spread complexity. The Krylov amplitudes are
\begin{eqnarray}
\varphi_n(t)
&\equiv&
\langle K_n|\psi(t)\rangle
=
\int d\epsilon\,\rho(\epsilon)e^{-\im \epsilon t}P_n(\epsilon)
\nonumber\\
&=&
\sqrt{\frac{(k)_n}{n!}}
\frac{(-\im \theta t)^n}
{(1+\im \theta t)^{n+k}} .
\end{eqnarray}
Their moduli squared again form a negative binomial distribution,
\be
|\varphi_n(t)|^2
=
\frac{(k)_n}{n!}
\frac{(\theta^2t^2)^n}
{(1+\theta^2t^2)^{n+k}}
=
\frac{(k)_n}{n!}
(1-p)^n p^k,
\qquad
p=\frac{1}{1+\theta^2t^2}.
\ee
The spread complexity and its growth rate are
\begin{eqnarray}
C(t)
&=&
\sum_{n=0}^{\infty}
n|\varphi_n(t)|^2
=
\frac{k(1-p)}{p}
=
k\theta^2t^2
=
\frac{M^2t^2}{8},
\nonumber\\
\dot C(t)
&=&
\frac{M^2}{4}t,
\qquad
\ddot C(t)
=
\frac{M^2}{4}.
\end{eqnarray}
Comparing with the width \eqref{wavewidth}, we find
\begin{eqnarray}
C(t)
&=&
\frac{1}{8}
\left[
M^2\sigma(t)^2-1
\right]
=
\frac{M^2}{8}
\left[
\sigma(t)^2-\sigma(0)^2
\right],
\\
\frac{dC(t)}{dt}
&=&
\frac{M^2}{8}
\frac{d(\sigma(t)^2)}{dt}.
\end{eqnarray}
Therefore, for the physical Hamiltonian, the packet center remains fixed, and spread complexity measures the dispersive broadening of the wave packet.

\section{Spread complexity in de Sitter space}
\label{sec_dS}
We now apply the same logic to Lorentzian dS$_2$. No detailed form of the dS/CFT correspondence will be assumed \cite{Strominger:2001pn,Witten:2001kn}. The relevant question is instead more elementary: given a particle Hilbert space in de Sitter space, can spread complexity be related to a classical observable on the particle phase space?

There is an immediate difference from AdS. In global dS$_2$ there is no globally timelike Killing vector, and the natural Hamiltonian is time-dependent. The standard time-independent Krylov construction is therefore not directly applicable. We restrict attention to the static patch, where a conserved energy exists. Quantization in this patch leads to a continuous spectrum and to a principal-series representation of $SL(2,\mathbb R)$, rather than the discrete series that appeared in AdS. Consequently there is no normalizable primary state analogous to the AdS state. This changes the relation between spread complexity and classical particle motion.

We first quantize the classical phase space of a massive particle in the static patch. We then discuss two kinds of initial states. A generic semiclassical wave packet in the energy representation has a well-defined particle interpretation, but its spread complexity is dominated by dispersion and does not coincide with a simple spacetime charge. By contrast, in the examples studied below, special states associated with the principal-series representation lead to Lanczos coefficients growing linearly with the Krylov index and to exponential late-time growth of spread complexity. In an appropriate boundary-localized limit, the resulting growth rate is proportional to the radial momentum of a static-patch particle.

\subsection{Quantization of a free massive particle in \texorpdfstring{dS$_2$}{dS2}}

Consider a free massive particle moving in Lorentzian dS$_2$, embedded in $\mathbb{R}^{1,2}$ with metric
\be
\eta_{AB}=\operatorname{diag}(-1,1,1),
\qquad
A,B=0,1,2,
\ee
as the hyperboloid
\be
-(X^0)^2+(X^1)^2+(X^2)^2=1 .
\ee
The embedding makes the $SO(1,2)$ isometry manifest. This group is locally isomorphic to $SL(2,\mathbb R)$. The geodesic equation is
\be
\frac{d^2X^A}{d\tau^2}-X^A=0,
\ee
where $\tau$ is proper time, normalized by
\be
\eta_{AB}\dot X^A\dot X^B=-1 .
\ee
The general timelike solution is
\be
X^A(\tau)=e_1^A\cosh\tau+e_2^A\sinh\tau ,
\ee
where
\be
\eta_{AB}e_1^Ae_1^B=1,
\qquad
\eta_{AB}e_2^Ae_2^B=-1,
\qquad
\eta_{AB}e_1^Ae_2^B=0 .
\ee
The presymplectic form on the space of solutions is
\be
\omega
=
m
\left(
de_1^0\wedge de_2^0
-
de_1^1\wedge de_2^1
-
de_1^2\wedge de_2^2
\right).
\label{presym2}
\ee
After imposing the constraints and quotienting by worldline reparametrizations, the physical phase space is two-dimensional. 

We use the following static-patch coordinates $(T,\chi)$,
\be
X^0=\sech\chi\sinh T,
\qquad
X^1=\sech\chi\cosh T,
\qquad
X^2=\tanh\chi ,
\ee
in which the metric takes the conformally flat form
\be
ds^2
=
\sech^2\chi
\left(
-dT^2+d\chi^2
\right).
\ee
A timelike geodesic in these coordinates can be written as
\be
\sinh\chi
=
\tanh\chi_0\,\sinh(T-T_0),
\qquad
\chi_0\in[0,\infty).
\label{dSstatic}
\ee
Here $T_0$ is the static time at which the geodesic passes through the center of the static patch, $\chi=0$, while $\chi_0$ controls its conserved energy.
To express the symplectic form in these variables, choose a gauge in which $\tau=0$ corresponds to $T=T_0$. The embedding data are then
\be
e_1^A
=
(\sinh T_0,\cosh T_0,0),
\ee
and
\be
e_2^A
=
(\cosh\chi_0\cosh T_0,
 \cosh\chi_0\sinh T_0,
 \sinh\chi_0).
\ee
Substitution into \eqref{presym2} gives the symplectic form
\be
\omega
=
m\sinh\chi_0\,dT_0\wedge d\chi_0 .\label{symds}
\ee
The conserved $SO(1,2)$ charges
\be
L_{AB}
=
m(X_A\dot X_B-X_B\dot X_A)
\ee
are
\be
L_{01}=m\cosh\chi_0,
\qquad
L_{02}=-m\sinh\chi_0\sinh T_0,
\qquad
L_{12}=m\sinh\chi_0\cosh T_0 .
\ee
In particular, the static-patch Hamiltonian is
\be
H_{\rm dS}=L_{01}=m\cosh\chi_0 .
\ee

The symplectic form \eqref{symds} is formally similar to the AdS result \eqref{sympar}. One might therefore also introduce the complex variable
\be
z
=
\tanh\frac{\chi_0}{2}\,e^{-\im T_0},
\ee
which gives
\be
\omega
=
m
\frac{2\im \,dz\wedge d\bar z}{(1-|z|^2)^2}.
\ee
This would lead to a discrete $SL(2,\mathbb R)$ representation. However, $T_0$ is not periodic in the Lorentzian static patch. The variable $z$ therefore compactifies the time direction and artificially discretizes the spectrum. Although one can recover the continuous spectrum by a suitable limiting procedure, it is cleaner to use a noncompact canonical polarization.

If we define
\be
E
=
m\cosh\chi_0
=
L_{01}
=
H_{\rm dS},\quad dE=m\sinh\chi_0\,d\chi_0,
\ee
then the symplectic form becomes
\be
\omega=dT_0\wedge dE,
\qquad
E\geq m .
\ee
This phase space, with coordinates $(T_0,E)$, is the same as the one appearing in pure JT gravity \cite{Harlow:2018tqv}. It is useful to subtract the rest energy,
\be
\epsilon=E-m\geq0 .
\ee
The particle Hilbert space is
\be
\mathcal H_{\rm dS}
=
L^2(\mathbb R_+,d\epsilon),
\ee
and the Hamiltonian acts by multiplication,
\be
\hat H_{\rm dS}\psi(E)=E\psi(E).
\ee

The same continuous spectrum appears in the Klein-Gordon quantization in the static patch \cite{Sun:2021thf}. The field expansion is
\be
\hat\phi(T,\chi)
=
\int_0^\infty d\omega\,
\left[
\hat a_\omega\Phi_\omega(T,\chi)
+
\hat a_\omega^\dagger\Phi_\omega^*(T,\chi)
\right],
\ee
with
\be
\Phi_\omega(T,\chi)
=
N_\omega
e^{-\im \omega T}
\cosh^{-1/2}\chi\,
P_{-\frac12+\im \omega}^{\nu}(\tanh\chi),
\qquad
\nu=\sqrt{m^2-\frac14},
\ee
and
\be
N_\omega
=
\sqrt{
\frac{\omega\tanh(\pi\omega)}{\pi}
}
\Gamma\left(\frac12+\nu+\im\omega\right).
\ee
Thus static-patch quantization gives a principal-series representation of $SL(2,\mathbb R)$, not a discrete-series representation. There is no normalizable primary state from which one can build the Krylov basis as in AdS. To our knowledge, spread complexity for continuous principal-series representations has not been systematically studied.

There is also a quantization-theoretic subtlety. The charges $L_{02}$ and $L_{12}$ contain $\sinh T_0$ and $\cosh T_0$, and hence involve all powers of $T_0$ in their Taylor expansions. They cannot be quantized without ambiguity by a naive canonical prescription. This is an instance of the obstruction expressed by the Groenewold-van Hove theorem \cite{groenewold12principles,van1951certaines}. One may try to promote $T_0$ to $-\im\partial_E$, but this introduces ordering ambiguities. Moreover, since $\epsilon$ is supported on $\mathbb R_+$, the operator $-\im\partial_E$ is not automatically self-adjoint; depending on the deficiency indices, it may have no self-adjoint extension, a unique extension, or infinitely many extensions \cite{Weyl:1910,Neumann1930}.

\subsection{Semiclassical wave packets in the energy representation}

Even without a canonical coherent state, one can choose a semiclassical wave packet in the energy representation and compute its spread complexity. Let
\be
|\Psi_0\rangle
=
\int_0^\infty d\epsilon\,c(\epsilon)|\epsilon\rangle,
\qquad
\hat H_{\rm dS}|\epsilon\rangle=(m+\epsilon)|\epsilon\rangle,
\qquad
\hat{T}_0|\epsilon\rangle=\im \frac{\partial}{\partial \epsilon}|\epsilon\rangle,
\ee
with spectral measure
\be
d\mu(\epsilon)=|c(\epsilon)|^2d\epsilon .
\ee
On the half-line, a convenient model for a semiclassical distribution is the gamma distribution
\be
d\mu(\epsilon)
=
\frac{\lambda^\beta}{\Gamma(\beta)}
\epsilon^{\beta-1}e^{-\lambda\epsilon}d\epsilon,
\qquad
\beta>0,
\qquad
\lambda>0 .
\label{dSgamma}
\ee
It has
\be
\langle\epsilon\rangle=\frac{\beta}{\lambda},
\qquad
{\rm Var}(\epsilon)=\frac{\beta}{\lambda^2}.
\ee
For large $\beta$ at fixed mean, the distribution becomes sharply peaked and is locally Gaussian near its saddle point. It therefore has the usual semiclassical wave-packet interpretation.

To describe a classical particle with static-patch energy
\be
E_0=m\cosh\chi_0,
\ee
we choose
\be
\frac{\beta}{\lambda}
=
m(\cosh\chi_0-1).
\label{dSclassicalenergy}
\ee
As in the flat-space analysis, the gamma distribution leads to Laguerre polynomials and gives
\be
C_{\rm dS}(T)
=
\frac{\beta}{\lambda^2}T^2.
\label{dScomplexity}
\ee
This result has a simple interpretation. The spread complexity is controlled by the variance of the energy distribution. It therefore measures the dispersive broadening of the wave packet in the variable conjugate to the energy, rather than the classical radial motion of the particle.

To compare this behavior with a classical observable, introduce the proper radial coordinate $\rho$ by
\be
\sin\rho=\tanh\chi,
\ee
so that the metric becomes
\be
ds^2=-\cos^2\rho\, dT^2+d\rho^2 .
\ee
For the center-crossing geodesic \eqref{dSstatic}, the canonical radial momentum is
\be
p_\rho(T)=m\sinh\chi_0\cosh(T-T_0),
\ee
which does not vanish at the center-crossing time $T=T_0$.

For comparison, one may instead consider a particle initially at rest at $\rho=\rho_0$, with the turning-point initial condition $p_\rho(0)=0$. The radial momentum then evolves as
\be
p_\rho(T)=m\sin\rho_0\sinh T .
\label{dsradial}
\ee
For this second family of geodesics, the conserved static-patch energy is $E=m\cos\rho_0$; the parameter $\rho_0$ should therefore not be identified with the parameter $\chi_0$ appearing in \eqref{dSstatic}. Both families exhibit exponential growth of the radial momentum at late static time.
Thus the classical radial momentum grows exponentially in static time, whereas \eqref{dScomplexity} grows only quadratically. The mismatch is not a paradox. The gamma wave packet is semiclassical in energy, but it is not a coherent state adapted to the $SL(2,\mathbb R)$ dynamics. Its spread complexity is governed by dispersion in the energy representation, not by rigid motion along a classical orbit.

\subsection{A coherent state in the principal series}

The preceding discussion used the energy representation. We now use the principal-series representation directly. This makes the $SL(2,\mathbb R)$ structure explicit and gives a class of solvable Krylov problems. Let $\{P,D,K\}$ denote the standard generators of $SL(2,\mathbb R)$, with commutation relations
\be
[D,P]=P,
\qquad
[D,K]=-K,
\qquad
[K,P]=2D.
\ee
Their relation to the generators used above is
\be
L_{02}=D,
\qquad
L_{21}=\frac{1}{2}(P+K),
\qquad
L_{01}=\frac{1}{2}(P-K).
\ee
Introduce the Hermitian operator
\be
L_0=\im L_{12}=-\frac{\im}{2}(P+K),
\ee
and the ladder operators
\be
L_{\pm}=-\frac{\im}{2}(P-K)\mp D .
\ee
They obey
\be
[L_0,L_\pm]=\pm L_{\pm},
\qquad
[L_-,L_+]=2L_0,
\qquad
L_-^\dagger=L_+ .
\ee
The static-patch Hamiltonian is\footnote{Indeed,
\be
[H_{\text{s}},\phi(t,\chi)]
=
-\im \partial_t \phi
=
\im \mathcal{L}_{01}\phi
=
[-\im L_{01},\phi(t,\chi)],
\ee
where
\be
\mathcal{L}_{AB}=X_A\partial_B-X_B\partial_A
\ee
are the Killing vector fields.}
\be
H_{\text{s}}
=
-\im L_{01}
=
\frac{L_++L_-}{2}.
\ee

The principal-series representation is
\be
L_0|n\rangle=n|n\rangle,
\qquad
L_\pm |n\rangle=(n\pm \Delta)|n\pm 1\rangle,
\qquad
\Delta=\frac{1}{2}+\im\mu,
\qquad
n\in \mathbb{Z},
\qquad
\mu\in\mathbb{R}.
\ee
This representation is realized by global one-particle excitations \cite{Sun:2021thf},
\be
|n\rangle=a_n^\dagger|\Omega\rangle,
\qquad
a_n|\Omega\rangle=0,
\ee
where $|\Omega\rangle$ is the Euclidean vacuum. In global coordinates,
\be
ds^2=\frac{-dt^2+d\varphi^2}{\cos^2 t},
\ee
the Klein-Gordon field has the expansion
\be
\phi(t,\varphi)
=
\sum_n
\phi_n a_n+\phi_n^* a_n^\dagger,
\qquad
\phi_n
=
\frac{e^{-\im n\varphi}}{\sqrt{2\pi}}g_n(t),
\ee
with
\be
g_n(t)
=
\frac{\Gamma(n+\bar{\Delta})}{\sqrt{2}}
e^{-\im n t}\,
{}_2\tilde{F}_1
\left(
\Delta,\bar{\Delta};n+1;
\frac{1}{1+\exp(2 \im t)}
\right),
\qquad
\bar{\Delta}=1-\Delta .
\ee
Note that a static-patch mode $\Phi_\omega(T,\chi)$ is defined only in one static patch, for instance the right patch. To relate it to the global modes $\phi_n(t,\varphi)$ one must include both right and left static modes.

A formal generalized coherent state in the principal series may be written as
\be
|\zeta\rangle
=
e^{\xi L_+-\bar{\xi}L_-}|0\rangle,
\qquad
\xi=|\xi|e^{\im \varphi},
\qquad
\zeta=\tanh|\xi|e^{\im \varphi}.
\ee
We first take the simplest state $|0\rangle$ as the initial state and compute its return amplitude,
\be
A_0(t)
=
\langle 0|
e^{-\im \frac{L_++L_-}{2}t}
|0\rangle .
\ee
In the circle realization of the principal series representation, where
\be
|n\rangle
\leftrightarrow
\frac{e^{\im n\theta}}{\sqrt{2\pi}},
\ee
the generators act as
\begin{eqnarray}
L_0&=&-\im\partial_\theta,
\nonumber\\
L_\pm&=&e^{\pm\im\theta}
\left(
-\im\partial_\theta\pm\Delta
\right).
\end{eqnarray}
The Hamiltonian becomes
\be
H_s=-\im \cos\theta\,\partial_\theta+\im \Delta\sin\theta .
\ee
For the initial wavefunction
\be
\psi_0(\theta)=\frac{1}{\sqrt{2\pi}},
\ee
the Schr\"odinger equation gives
\be
\psi(t,\theta)
=
e^{-\im H_s t}\psi_0(\theta)
=
\frac{1}{\sqrt{2\pi}}
\left(
\cosh t-\sinh t \sin\theta
\right)^{-\Delta}.
\ee
The same expression follows from the group action. Let
\be
g_t=e^{-\im H_s t}.
\ee
In the circle realization, $g_t$ acts on functions $f(z=e^{\im\theta})$ by \cite{Perelomov:1986uhd}
\be
T(g_t)f(z)
=
|\beta z+\bar{\alpha}|^{-\Delta}f(z_g),
\qquad
z_g
=
\frac{\alpha z+\bar{\beta}}{\beta z+\bar{\alpha}},
\ee
where
\be
g_t
=
\frac{1}{\sqrt{1-|\zeta|^2}}
\begin{pmatrix}
1&-\zeta\\
-\bar{\zeta}&1
\end{pmatrix}
=
\begin{pmatrix}
\alpha&\beta\\
\bar{\beta}&\bar{\alpha}
\end{pmatrix},
\qquad
\zeta=-\im \tanh\frac{t}{2}.
\ee
Thus
\be
A_0(t)
=
\frac{1}{2\pi}
\int_0^{2\pi}
\left(
\cosh t-\sinh t \sin\theta
\right)^{-\Delta}
d\theta
=
P_{\Delta-1}(\cosh t),
\ee
where $P_\nu(x)$ is the Legendre function.

Using the spectral representation
\be
A_0(t)=\int dE\,e^{-\im Et}\rho(E),
\ee
we obtain the spectral density by Fourier inversion:
\be
\rho(E)
=
\frac{1}{2\pi}
\int_{-\infty}^{\infty}dt\,e^{\im Et}A_0(t)
=
\frac{\cosh(\pi \mu)}{4\pi^3}
\left|
\Gamma\left(\frac{\im(\mu+E)}{2}+\frac{1}{4}\right)
\Gamma\left(\frac{\im(E-\mu)}{2}+\frac{1}{4}\right)
\right|^2 ,
\ee
where we have used Eq.~(7.165) of \cite{Gradshteyn:2014tis}.
The orthogonal polynomials associated with this spectral measure are continuous Hahn polynomials. We use the convention in which their orthogonality relation is
\begin{eqnarray}
&&
\frac{1}{2\pi}
\int_{-\infty}^\infty
\Gamma(a+\im x)\Gamma(b+\im x)
\Gamma(c-\im x)\Gamma(d-\im x)
p_n(x;a,b,c,d)p_m(x;a,b,c,d)\,dx
\nonumber\\
&&\hspace{1cm}
=
\frac{
\Gamma(a+c+n)\Gamma(a+d+n)\Gamma(b+c+n)\Gamma(b+d+n)
}{
\Gamma(n+1)(a+b+c+d+2 n-1)\Gamma(a+b+c+d+n-1)
}
\delta_{nm}.
\end{eqnarray}
In the present case, the parameters are
\be
a=d=\frac{1}{4}+\frac{\im\mu}{2},
\qquad
b=c=\frac{1}{4}-\frac{\im\mu}{2},
\qquad
x=\frac{E}{2}.
\ee
With this identification, the functions
\be
\psi_n(E)
=
\frac{\pi}{\sqrt{\cosh \pi\mu}}
\frac{\Gamma(n+1)}{\Gamma(n+\frac{1}{2})}
\sqrt{
\frac{2}{
\Gamma(n+\frac{1}{2}+\im \mu)
\Gamma(n+\frac{1}{2}-\im \mu)
}
}
p_n\left(\frac{E}{2};a,b,c,d\right)
\ee
are orthonormal with respect to the spectral density,
\be
\int_{-\infty}^\infty \rho(E)\psi_n(E)\psi_m(E)\,dE
=
\delta_{nm}.
\ee
The recurrence relation gives the Lanczos coefficients
\be
a_n=0,
\qquad
b_n=\frac{1}{2}\sqrt{\left(n-\frac{1}{2}\right)^2+\mu^2},
\qquad
n\geq2,
\qquad
b_1=\sqrt{\frac{1}{2}\left(\mu^2+\frac{1}{4}\right)} .
\ee
Solvable Krylov problems based on continuous Hahn polynomials were introduced in \cite{Gamayun:2025hvu}. The construction above gives such a model a direct physical realization in terms of a principal-series coherent state. The Krylov amplitudes $\varphi_n(t)$ can be written in closed form in terms of a ${}_3F_2$ hypergeometric function \cite{Gamayun:2025hvu}. The resulting expression for the spread complexity is not particularly illuminating, but its asymptotic behavior is simple. At early times,
\be
C(t)
=
b_1^2 t^2+\mathcal{O}(t^4)
\sim
\frac{\Delta(1-\Delta)}{2}t^2 .
\ee
At large Krylov index, the Lanczos coefficients behave as
\be
b_n\sim \frac{n}{2}.
\ee
The corresponding Krylov dynamics is therefore expected to exhibit exponential spreading, and the late-time complexity behaves as
\be
C(t)\sim e^t .
\ee
This agrees qualitatively with the exponential growth of the classical radial momentum in \eqref{dsradial}.

There is, however, an important caveat. For the state $|0\rangle$, all one-point functions of the generators vanish:
\be
\langle 0|L_0|0\rangle
=
\langle 0|L_\pm|0\rangle
=
0 .
\ee
Thus it is not clear that this state represents a single massive particle in one static patch. Indeed, the spectral density is symmetric,
\be
\rho(E)=\rho(-E).
\ee
In the large-$\mu$ limit, it approaches two sharp peaks,
\be
\rho(E)
\to 
\frac{1}{2}\delta(E+\mu)+\frac{1}{2}\delta(E-\mu),
\ee
suggesting a left-right particle pair rather than a single-particle state in one patch. We therefore consider a different class of states, motivated by the extrapolation dictionary \cite{Strominger:2001pn}.

Near the future boundary, the scalar field behaves as \cite{Strominger:2001pn}
\be
\phi\left(t=\frac{\pi}{2}-\delta,\varphi\right)
\sim
\delta^\Delta \mathcal{O}_\Delta(\varphi)
+
\delta^{\bar{\Delta}}\mathcal{O}_{\bar{\Delta}}(\varphi),
\ee
where $\mathcal{O}_\Delta(\varphi)$ and $\mathcal{O}_{\bar{\Delta}}(\varphi)$ are boundary primary operators. In particular,
\be
\co_\Delta(\varphi)
=
\sum_n
\left[
(-\im)^{n+\Delta}
\frac{e^{-\im n\varphi}}{\sqrt{2\pi}}a_n
+
\im^{n+\Delta}
\frac{\Gamma(n+\Delta)}{\Gamma(n+\bar{\Delta})}
\frac{e^{\im n\varphi}}{\sqrt{2\pi}}a_n^\dagger
\right].
\ee
Mimicking the AdS construction, one may consider
\be
|\Delta,\varphi\rangle
=
\co_\Delta(\varphi)|\Omega\rangle
=
\sum_n
\im^{n+\Delta}
\frac{\Gamma(n+\Delta)}{\Gamma(n+1-\Delta)}
\frac{e^{\im n \varphi}}{\sqrt{2\pi}}|n\rangle .
\ee
These states form a continuous basis of the representation, although they are not normalizable. In the large-$\mu$ limit, for fixed $n$, Stirling's formula gives
\be
\frac{\Gamma(n+\Delta)}{\Gamma(n+1-\Delta)}
=
\frac{\Gamma(n+\frac12+\im\mu)}
{\Gamma(n+\frac12-\im\mu)}
\sim
e^{2\im\mu(\ln\mu-1)}
e^{\im n\pi}.
\ee
Therefore, up to an overall phase independent of $n$, one finds
\be
\langle\theta|\Delta,\varphi\rangle
\sim
\frac{1}{2\pi}
\sum_n
e^{\im n(\frac{\pi}{2}+\pi+\varphi+\theta)}
\sim
\delta_{2\pi}\left(\varphi+\theta-\frac{\pi}{2}\right),
\label{deltapeak}
\ee
where $\delta_{2\pi}$ denotes the periodic delta function on the circle.
 This suggests a natural regularization,
\be
|\psi_{\beta,\varphi}\rangle
=
\mathcal{N}_\beta
e^{-\beta |L_0|}
\mathcal{O}_\Delta(\varphi)|\Omega\rangle,
\qquad
\mathcal{N}_\beta=\sqrt{2\pi \tanh \beta}.
\ee
This is the dS analogue of the regulated AdS state \eqref{adsstate}. For $\mu\gg1$, its wavefunction in the circle realization is
\be
\langle \theta|\psi_{\beta,\varphi}\rangle
=
\frac{\mathcal{N}_\beta}{2\pi}
\sum_n
e^{\im n(\varphi+\theta-\frac{\pi}{2})}
e^{-\beta |n|}
=
\frac{\mathcal{N}_\beta}{2\pi}
\frac{\sinh\beta}
{\cosh\beta-\cos(\varphi+\theta-\frac{\pi}{2})}.
\ee
Under static-patch time evolution, the wavefunction transforms as
\be
e^{-\im t H_s}
\langle \theta|\psi_{\beta,\varphi}\rangle
=
\left(
\cosh t-\sinh t \sin\theta
\right)^{-\Delta}
\langle \theta_t|\psi_{\beta,\varphi}\rangle,
\ee
with
\be
\tan\left(\frac{\pi}{4}+\frac{\theta_t}{2}\right)
=
e^{-t}
\tan\left(\frac{\pi}{4}+\frac{\theta}{2}\right).
\ee
The return amplitude is therefore
\be
A_0(t)
=
\int d\theta\,
\langle\theta|\psi_{\beta,\varphi}\rangle^*
\left(
\cosh t-\sinh t \sin\theta
\right)^{-\Delta}
\langle \theta_t|\psi_{\beta,\varphi}\rangle .
\ee
For generic values of $\beta$, this integral is not expressible in elementary functions. The limit $\beta\to0$ is tractable. In this limit, the wavefunction becomes sharply localized, and a nonvanishing return amplitude requires the localization point to be fixed under the flow,
\be
\theta=\pm \frac{\pi}{2}.
\ee
By \eqref{deltapeak}, this corresponds to
\be
\varphi=0,\pi .
\ee
It is useful to trivialize the flow by introducing
\be
x=\tan\left(\frac{\pi}{4}+\frac{\theta}{2}\right),
\ee
such that
\be
x_t=e^{-t}x,
\qquad
d\theta=\frac{2}{1+x^2}dx.
\ee
Let us first consider the fixed point
\be
\theta=-\frac{\pi}{2},
\ee
which, by \eqref{deltapeak}, corresponds to $\varphi=\pi$. At this point,
\be
\left(
\cosh t-\sinh t \sin\theta
\right)^{-\Delta}
=
e^{-\Delta t}.
\ee
In the sharply localized limit $\beta\ll 1$, the wavefunctions near this fixed point take the approximate form
\begin{eqnarray}
\langle \theta|\psi_{\beta,\varphi}\rangle
&\approx&
\sqrt{\frac{\beta}{2\pi}}\,
\frac{2 \beta \left(x^2+1\right)}
{\left(\beta^2+4\right)x^2+\beta^2},
\nonumber\\
\langle \theta_t|\psi_{\beta,\varphi}\rangle
&\approx&
\sqrt{\frac{\beta}{2\pi}}\,
\frac{2 \beta \left(x^2 e^{-2t}+1\right)}
{\left(\beta^2+4\right)x^2 e^{-2t}+\beta^2}.
\end{eqnarray}
The overlap integral can then be evaluated analytically. Keeping only the leading term in $\beta$, one obtains
\be
A_0^-(t)
=
e^{-\Delta t}
\frac{2e^t}{1+e^t}
=
\frac{e^{-\im\mu t}}{\cosh(t/2)} .
\ee
The other fixed point,
\be
\theta=\frac{\pi}{2},
\ee
corresponds to $\varphi=0$. Repeating the same analysis gives
\be
A_0^+(t)
=
e^{\Delta t}
\frac{2}{1+e^t}
=
\frac{e^{\im\mu t}}{\cosh(t/2)} .
\ee
Fourier transforming these amplitudes gives the corresponding spectral densities,
\begin{eqnarray}
\rho^\pm(E)
&=&
\frac{1}{2\pi}
\int_{-\infty}^{\infty}dt\,
e^{\im Et}
\frac{e^{\pm \im\mu t}}{\cosh(t/2)}
\nonumber\\
&=&
\sech\!\left[\pi(E\pm\mu)\right].
\end{eqnarray}
Each density has a single peak,
\be
\rho^+(E):\quad E=-\mu,
\qquad
\rho^-(E):\quad E=+\mu,
\ee
suggesting that these states may describe particles localized in opposite static patches. In what follows, we focus on one branch, for instance
\be
\rho(E)\equiv \rho^+(E)
=
\sech\!\left[\pi(E+\mu)\right].
\ee

The orthogonal polynomials associated with this weight are shifted Meixner-Pollaczek polynomials,
\be
p_n(E)
=
P_n^{(1/2)}
\left(E+\mu;\frac{\pi}{2}\right),
\ee
where
\be
P_n^{(\lambda)}(x;\phi)
=
\frac{(2\lambda)_n}{n!}
e^{\im n\phi}
{}_2F_1
\left(
-n,\lambda+\im x;2\lambda;1-e^{-2\im\phi}
\right).
\ee
The recurrence relation
\be
2x\sin\phi\,P_n^{(\lambda)}(x;\phi)
=
(n+1)P_{n+1}^{(\lambda)}(x;\phi)
-
2(n+\lambda)\cos\phi\,P_n^{(\lambda)}(x;\phi)
+
(n+2\lambda-1)P_{n-1}^{(\lambda)}(x;\phi)
\ee
gives
\be
a_n=-\mu,
\qquad
b_n=\frac{n}{2}.
\ee
The constant shift $a_n=-\mu$ contributes only an overall phase. The Krylov dynamics is therefore governed by the linear Lanczos growth $b_n=n/2$, and the spread complexity is
\be
C(t)
=
\sinh^2\frac{t}{2}.
\ee
Consequently
\be
\frac{dC(t)}{dt}
=
\frac{1}{2}\sinh t .
\ee
Upon identifying $t$ with the static time $T$, this has the same time dependence as the classical radial momentum \eqref{dsradial}:
\be
\frac{dC(t)}{dt}
\propto
p_\rho(T).
\ee
Thus, if the boundary-localized states considered here select a single static patch, spread complexity again admits a direct classical interpretation. The contrast with the gamma wave packet is useful: a semiclassical energy distribution alone is not enough. The state must also be adapted to the symmetry and to the relevant classical orbit.
\subsection{Comments on spread complexity in JT gravity and DSSYK}

Krylov observables in JT gravity and DSSYK \cite{Jackiw:1984je,Teitelboim:1983ux,Mertens:2017mtv} provide several related, but distinct, bulk constructions. For appropriate two-sided Hartle--Hawking states and in suitable continuum or semiclassical regimes, spread complexity can be identified with the renormalized length of the wormhole connecting the two boundaries \cite{Lin:2022rbf,Rabinovici:2023yex,Aguilar-Gutierrez:2025pqp}. Related results for operator K-complexity, including setups with matter, have been obtained in \cite{Jian:2020qpp,Ambrosini:2024sre,Aguilar-Gutierrez:2025pqp}. These operator and state complexities should be distinguished. In particular, the relation between operator complexity and proper radial momentum discussed in \cite{Aguilar-Gutierrez:2025pqp} does not automatically imply the same relation for spread complexity with a matter insertion.

The initial states used in these constructions should be specified separately. The chord-basis analyses of \cite{Lin:2022rbf,Rabinovici:2023yex} start from the zero-chord state, equivalently the infinite-temperature thermofield double state in the effective DSSYK description. By contrast, \cite{Aguilar-Gutierrez:2025pqp} considers the two-sided Hartle--Hawking state prepared by the DSSYK path integral, with matter insertions incorporated through the path-integral construction. The analysis of \cite{Fu:2025kkh} instead uses coherent states obtained by acting with the $q$-deformed displacement operator on the zero-chord state, while \cite{Aguilar-Gutierrez:2026nmd} discusses generalized $q$-coherent ETW-brane states in deformed DSSYK models. These examples illustrate that the classical bulk interpretation of complexity depends on the initial state: the state must be specified, and different choices can lead to different classical limits and bulk descriptions.

There is also an important distinction between DSSYK/JT duality and the probe analysis in the present paper. In the continuum JT regime, DSSYK describes the gravitational dynamics itself, rather than a probe theory in a fixed gravitational background. The dictionary used here for a local bulk probe therefore cannot be imported directly. Quantization of JT gravity leads to representations of $SL(2,\mathbb{R})$, more precisely $SL^+(2,\mathbb{R})$ \cite{Blommaert:2018oro}. Evolution of a generalized coherent state in such a representation can produce exponentially growing spread complexity, whereas the semiclassical wormhole length in the DSSYK construction grows linearly. This difference illustrates the dependence of Krylov complexity on the state, Hamiltonian, and representation, rather than a contradiction between the two constructions.

Finally, nonperturbative features of the Lanczos spectrum and the associated saturation of complexity have been studied using finite-dimensional Hamiltonians modeling the DSSYK density of states and random-matrix constructions \cite{Nandy:2024tridiagonal,Balasubramanian:2024lqk}. It would be useful to understand how these effects appear in a directly quantized bulk description. Sine-dilaton gravity provides a promising setting for this purpose \cite{Blommaert:2024ymv,Blommaert:2023wad,Blommaert:2023opb}, especially given the match between DSSYK spread complexity and wormhole length including quantum corrections discussed in \cite{Heller:2024ldz}.

\section{Summary and Outlook}
\label{sec_summary}
In this section we first summarize our comments on the questions raised in Section \ref{sec_intro}, and then mention several directions for future work.

Spread complexity depends both on the initial, or reference, state and on the Hamiltonian. Its holographic interpretation should therefore depend on both data. In particular, the proposal of \cite{Caputa:2024sux}, which identifies the growth rate of spread complexity with the proper radial momentum, should be understood as a statement for a particular class of states. To compute a holographic spread complexity within this proposal, one has to specify the quantum state.

As stressed in \cite{Li:2025fqz}, momentum is not a coordinate-dependent quantity. Thus the essential point of this proposal is not which momentum is the correct one, but rather that a quantum observable can admit a classical interpretation, in analogy with the geometric interpretation of entanglement entropy. The main comment of the present work is that taking only the semiclassical limit does not guarantee such a classical interpretation. In the examples studied here, a sufficient mechanism is to choose an initial state adapted to the symmetry, such as a generalized coherent state, together with a Hamiltonian belonging to the corresponding symmetry algebra. The evolution then preserves the coherent-state manifold and avoids generating dispersion. This mechanism is sufficient for the examples considered here, but is not intended to exclude other classes of initial states with classical limits.

To support these arguments, we computed the spread complexity directly in the bulk theory. The boundary theory is used to motivate the initial state and, through the extrapolation dictionary, to map it to a bulk one-particle state. Once this state is specified, however, the subsequent Krylov construction depends only on the bulk Hilbert space and Hamiltonian. In this sense, our analysis provides a bulk derivation of the proposal \cite{Caputa:2024sux} in the regime considered here. The boundary theory therefore plays no further essential role in the computational steps after the initial state has been fixed. By directly quantizing a free particle in AdS space, we also realize a quantization of the complexity variable as spread complexity in the CAny proposal. At the quantum level, these complexity observables are physical observables on the Hilbert space. When they do not commute with the Hamiltonian, they can diagnose the dynamics induced by the Hamiltonian on a chosen state.

Here, a typical state is one whose dynamics reflect universal properties of the Hamiltonian or of the system. By construction, spread complexity is the simplest complexity observable with no further input beyond the Hamiltonian and the chosen reference state. Since it can be computed directly in the bulk theory, one can examine the proposal beyond AdS spacetime or with other probes, for example, a non-free particle.

In flat spacetime, a coherent state can still be defined, but the physical Hamiltonian does not belong to the corresponding algebra. As a result, the growth of spread complexity comes purely from dispersion, since the Hamiltonian commutes with the momentum. In de Sitter space, we construct two types of coherent states. Since in this case the Hamiltonian belongs to the algebra, we find that the growth rate of spread complexity matches the proper radial momentum, up to an overall factor.

Our point of view raises a question about typical states. If spread complexity is to capture the universal dynamics of the system, one should understand why generalized coherent states, which look rather special, are the appropriate states to consider. Generalized coherent states can be defined for many Lie groups once a representation and a suitable reference state have been specified. In the context of Krylov complexity, however, most works have been restricted to low-rank Lie groups, such as $SU(2)$, the Heisenberg group, and $SU(1,1)$. For higher-rank Lie groups, there exist inequivalent classes of coherent states. In an upcoming work, we find, already in the simplest example of the $SU(3)$ group, that spread complexity has new features, which deserve further study. Related $SU(N)$ spin systems, their de Sitter interpretation, and the associated Krylov observables have been studied in \cite{Aguilar-Gutierrez:2025quasinormal}.

Besides this, a natural extension is to consider more general probes, in the spirit of \cite{Balasubramanian:1998de} and of the CAny proposal. A probe defines a phase space that provides a reduced description of the spacetime dynamics. By quantizing the probe, one can obtain a more precise identification of quantum complexity in the boundary and bulk theories. This may shed light on a holographic interpretation of other quantum complexities. Recent progress along this direction has been made in \cite{Chatzis:2026ekd}.

Since the spread complexity can be computed directly in the bulk theory at the quantum level, and since in principle the construction does not rely on exact symmetry, it is possible to consider more general asymptotically AdS backgrounds. Examples include conical AdS spacetime and shockwave AdS spacetime, backreacted respectively by massive and massless particles. Because the backreacted geometry does not have the exact global isometries, we expect the spread complexity of an approximate coherent state to receive contributions from both coherent motion and dispersive spreading. It would therefore not generally reduce to a single momentum observable. It would be interesting to study in detail whether spread complexity can tell us more about the gravity theory, rather than merely encoding information about symmetries.

It would also be interesting to generalize our discussion from global AdS to black hole backgrounds. In that case, the spread complexity becomes a concrete complexity observable in the CAny proposal, and may be useful for studying black hole physics. For example, recently in \cite{Miyaji:2025jxy}, complexity observables have been used to study aspects of the black hole interior.

\section*{Acknowledgments}

We are grateful to Chen Bai, Wen-Xin Lai, Weibo Mao, Farzad Omidi, Meng-Ting Wang, and Yu-Xuan Zhang for insightful discussions and valuable perspectives. JT is especially grateful to Sergio E. Aguilar-Gutierrez for his detailed comments on an earlier version of this manuscript. We also thank QIQG 2026 at Tsinghua University for its hospitality, where the final part of this work was completed. We acknowledge the use of K3.1 Pro (Kimi) to edit the manuscript and correct typographical errors.
\appendix
\section{Free scalar field on global \texorpdfstring{AdS$_{d+1}$}{AdS(d+1)}}
\label{appendix_adsd}

We give a bulk derivation of the spectral weights for the higher-dimensional analogue of the state \eqref{adsstate}. The result is again the negative binomial distribution \eqref{weight}.

Following \cite{Balasubramanian:1998sn,Kaplan:2016}, consider a scalar field on global AdS$_{d+1}$,
\be
ds^2
=
\frac{1}{\cos^2\rho}
\left(
-dt^2+d\rho^2+\sin^2\rho\,d\Omega_{d-1}^2
\right),
\qquad
0\leq \rho<\frac{\pi}{2}.
\ee
For a generic mass, the field admits the mode expansion
\begin{eqnarray}
\phi
=
\sum_{n=0}^{\infty}
\sum_{\ell=0}^{\infty}
\sum_J
\left(
a_{n\ell J}u_{n\ell J}
+
a_{n\ell J}^{\dagger}u_{n\ell J}^{*}
\right),
\end{eqnarray}
where
\begin{eqnarray}
u_{n\ell J}
&=&
\frac{1}{N_{\Delta n\ell}}\,
e^{-\im \omega_{n\ell}t}
Y_{\ell J}(\Omega)
\sin^\ell\rho\,\cos^\Delta\rho
\nonumber\\
&&\times
{}_2F_1
\left(
-n,\Delta+\ell+n;\ell+\frac d2;\sin^2\rho
\right).
\label{def_un_highd}
\end{eqnarray}
Here $Y_{\ell J}(\Omega)$ are scalar spherical harmonics on $S^{d-1}$, which are normalized as
\be
\int_{S^{d-1}}d\Omega\,
Y_{\ell J}^{*}(\Omega)Y_{\ell' J'}(\Omega)
=
\delta_{\ell\ell'}\delta_{JJ'} ,
\label{spherenormal}
\ee
The index $J$ denotes the remaining angular momentum quantum numbers. The conformal dimension and normal-mode frequencies are
\be
\Delta
=
\frac d2+\sqrt{\frac{d^2}{4}+m^2},
\qquad
\omega_{n\ell}
=
\Delta+\ell+2n .
\ee
With this convention, the normalization factor is given by \cite{Kaplan:2016}
\be
N_{\Delta n \ell}
=
(-1)^n
\sqrt{
\frac{
n!\,\Gamma^2\left(\ell+\frac d2\right)
\Gamma\left(\Delta+n-\frac{d-2}{2}\right)
}{
\Gamma\left(n+\ell+\frac d2\right)
\Gamma(\Delta+n+\ell)
}
}.
\ee
Using the extrapolation dictionary 
\be
\mathcal{O}(t,\Omega)
=
\lim_{\rho\to\frac{\pi}{2}}
(\cos\rho)^{-\Delta}
\hat \phi(t,\rho,\Omega)
\ee
and the identity
\be
{}_2F_1
\left(
-n,\Delta+\ell+n;\ell+\frac d2;1
\right)
=
(-1)^n
\frac{
\left(\Delta-\frac d2+1\right)_n
}{
\left(\ell+\frac d2\right)_n
},
\ee
one obtains the boundary mode expansion
\begin{eqnarray}
\mathcal{O}(t,\Omega)
=
\sum_{n,\ell,J}
\left(
\beta_{n\ell}
e^{-\im\omega_{n\ell}t}
Y_{\ell J}(\Omega)\hat a_{n\ell J}
+
\beta_{n\ell}^{*}
e^{\im\omega_{n\ell}t}
Y_{\ell J}^{*}(\Omega)\hat a_{n\ell J}^{\dagger}
\right),
\label{bme_highd}
\end{eqnarray}
with
\be
\beta_{n\ell}
=
\frac{(-1)^n}{N_{\Delta n\ell}}\,
\frac{
\left(\Delta-\frac d2+1\right)_n
}{
\left(\ell+\frac d2\right)_n
}.
\label{beta_highd}
\ee
Consider the Euclidean-regulated local excitation 
\be
|\psi_0\rangle
=
\frac{1}{\mathcal{N}_{\co}}
\co(z_0,\Omega_0)|\Omega\rangle ,
\qquad
0<z_0<1 .
\ee
It is equivalent to the bulk one-particle state 
\begin{eqnarray}
|\Psi_0\rangle
&=&
\frac{1}{\mathcal{N}_{\co}}
\sum_{n,\ell,J}
\beta_{n\ell}^{*}
z_0^{\ell+2n}
Y_{\ell J}^{*}(\Omega_0)
\hat a_{n\ell J}^{\dagger}
|\Omega\rangle_{\mathrm{bulk}}
\nonumber\\
&=&
\sum_{n,\ell,J}
c_{n\ell J}
\hat a_{n\ell J}^{\dagger}
|\Omega\rangle_{\mathrm{bulk}},
\label{bulkinitial_highd}
\end{eqnarray}
where
\be
c_{n\ell J}
=
\frac{1}{\mathcal{N}_{\co}}\,
\beta_{n\ell}^{*}
z_0^{\ell+2n}
Y_{\ell J}^{*}(\Omega_0).
\label{coef_highd}
\ee
Here $z_0$ is the radial Euclidean time variable. The probability carried by a single mode is
\be
w_{n\ell J}
=
|c_{n\ell J}|^2
=
\frac{1}{|\mathcal{N}_{\co}|^2}
|\beta_{n\ell}|^2
z_0^{2\ell+4n}
|Y_{\ell J}(\Omega_0)|^2 .
\ee
The spectral measure is therefore
\be
d\mu(E)
=
\sum_{n,\ell,J}
w_{n\ell J}
\delta(E-\Delta-\ell-2n)dE
=
\sum_{k=0}^{\infty}
W_k\delta(E-\Delta-k)dE ,
\label{spectral_measure_highd}
\ee
where
\be
k=\ell+2n,
\qquad
q=z_0^2,
\qquad
W_k
=
\frac{q^k}{|\mathcal{N}_{\co}|^2}
\sum_{\ell+2n=k}
|\beta_{n\ell}|^2
\sum_J
|Y_{\ell J}(\Omega_0)|^2 .
\label{Wk_pre}
\ee
To determine $W_k$, we introduce the generating function
\begin{eqnarray}
F(q)
&=&
\sum_{k=0}^{\infty}
q^k
\sum_{\ell+2n=k}
|\beta_{n\ell}|^2
\sum_J
|Y_{\ell J}(\Omega_0)|^2
\nonumber\\
&=&
\sum_{n,\ell,J}
q^{\ell+2n}
|\beta_{n\ell}|^2
|Y_{\ell J}(\Omega_0)|^2 .
\end{eqnarray}
The addition theorem for scalar spherical harmonics gives
\be
\sum_J
|Y_{\ell J}(\Omega_0)|^2
=
\frac{g_\ell}{\mathrm{Vol}(S^{d-1})},
\qquad
\mathrm{Vol}(S^{d-1})
=
\frac{2\pi^{d/2}}{\Gamma\left(\frac d2\right)} .
\ee
With the normalization \eqref{spherenormal}, the degeneracy of scalar harmonics on $S^{d-1}$ is \cite{vilenkin1978special}
\be
g_\ell
=
\frac{(2\ell+d-2)(\ell+d-3)!}{\ell!(d-2)!}
=
\frac{(2\ell+d-2)\Gamma(\ell+d-2)}
{\Gamma(\ell+1)\Gamma(d-1)} .
\ee
Substitution gives
\begin{eqnarray}
F(q)
&=&
\sum_{\ell=0}^{\infty}
\frac{g_\ell}{\mathrm{Vol}(S^{d-1})}
\sum_{n=0}^{\infty}
\frac{
q^{\ell+2 n}
\Gamma\left(-\frac d2+n+\Delta+1\right)
\Gamma(\ell+n+\Delta)
}{
\Gamma(n+1)
\Gamma\left(-\frac d2+\Delta+1\right)^2
\Gamma\left(\frac d2+\ell+n\right)
}
\nonumber\\
&=&
\sum_{\ell=0}^{\infty}
\frac{g_\ell}{\mathrm{Vol}(S^{d-1})}
\frac{
q^\ell
\Gamma(\ell+\Delta)
\,{}_2F_1
\left(
-\frac d2+\Delta+1,\ell+\Delta;\frac d2+\ell;q^2
\right)
}{
\Gamma\left(-\frac d2+\Delta+1\right)
\Gamma\left(\frac d2+\ell\right)
}
\nonumber\\
&=&
\sum_{\ell=0}^{\infty}
\frac{
2^{1-d}
\pi^{\frac12-\frac d2}
(d+2\ell-2)
q^\ell
\Gamma(d+\ell-2)
\Gamma(\ell+\Delta)
}{
\Gamma\left(\frac{d-1}{2}\right)
\Gamma(\ell+1)
\Gamma\left(-\frac d2+\Delta+1\right)
\Gamma\left(\frac d2+\ell\right)
}
\nonumber\\
&&\hspace{2.0cm}\times
{}_2F_1
\left(
-\frac d2+\Delta+1,\ell+\Delta;\frac d2+\ell;q^2
\right).
\label{Fq_sum_highd}
\end{eqnarray}
This sum can be evaluated by the generalized Gegenbauer generating function \cite{cohl2013generalization}
\begin{eqnarray}
\frac{1}{(1+\rho^2-2\rho x)^\nu}
&=&
\frac{\Gamma(\mu)}{\Gamma(\nu)}
\sum_{k=0}^{\infty}
(\mu+k) C_k^\mu(x)\rho^k
\frac{\Gamma(\nu+k)}{\Gamma(\mu+k+1)}
\nonumber\\
&&\hspace{2.0cm}\times
{}_2F_1
\left(
\nu+k,\nu-\mu;\mu+k+1;\rho^2
\right).
\label{gegenbauer_generalized}
\end{eqnarray}
with
\be
\mu=\frac d2-1,
\quad
\nu=\Delta,
\quad
\rho=q,
\quad
x=1,\quad
C_k^\mu(1)
=
\frac{\Gamma(k+2\mu)}
{\Gamma(k+1)\Gamma(2\mu)}.
\ee
It gives
\be
F(q)
=
\frac{\pi^{-d/2}\Gamma(\Delta)}
{2\Gamma\left(1-\frac d2+\Delta\right)}
(1-q)^{-2\Delta}.
\label{Fq_closed_highd}
\ee
Expanding \eqref{Fq_closed_highd} in powers of $q$,
\be
(1-q)^{-2\Delta}
=
\sum_{k=0}^{\infty}
\frac{(2\Delta)_k}{k!}q^k
=
\sum_{k=0}^{\infty}
(-1)^k
\binom{-2\Delta}{k}
q^k ,
\ee
one finds
\be
W_k
=
\frac{q^k}{|\mathcal{N}_{\co}|^2}
\frac{
\pi^{-\frac d2}
(-1)^k
\Gamma(\Delta)
\binom{-2\Delta}{k}
}{
2\Gamma\left(-\frac d2+\Delta+1\right)
}.
\ee
The normalization condition $\sum_{k=0}^{\infty}W_k=1$ fixes
\be
|\mathcal{N}_{\co}|^2
=
\frac{
\Gamma(\Delta)
}{
2\pi^{d/2}
\Gamma\left(1-\frac d2+\Delta\right)
}
(1-q)^{-2\Delta}.
\ee
Hence
\be
W_k
=
q^k(1-q)^{2\Delta}
\frac{(2\Delta)_k}{k!},
\qquad
W_k\sim \mathrm{NB}(2\Delta,1-q),
\ee
in agreement with \eqref{weight}.

\bibliographystyle{unsrturl}
\bibliography{HKref1}
\end{document}